\documentclass[
superscriptaddress,
nofootinbib,
amsmath,amssymb,
aps,
prd,
floatfix,
]{revtex4}

\usepackage{amsmath,amsfonts,amssymb,amsthm}
\usepackage{graphicx}
\usepackage{float}
\usepackage[latin1, utf8]{inputenc}
\usepackage{hyperref}
\usepackage{dsfont}
\usepackage{subcaption}
\usepackage{color}
\usepackage{multirow}  
\usepackage{empheq}
\usepackage{algorithm}
\usepackage{algpseudocode}
\usepackage{svg}

\newcommand{\nn}{\nonumber}

\newcommand{\Wi}{W \left(   j , i_n \right)}

\newcommand{\Nf}{\frac{1}{Z}}

\newcommand{\Wthree}[6]{\left(\begin{array}{ccc} #1 & #2 & #3 \\ #4 & #5 & #6 \end{array}\right)}
\newcommand{\Wfour}[9]{\left(\begin{array}{cccc} #1 & #2 & #3 & #4 \\ #5 & #6 & #7 & #8 \end{array}\right)^{(#9)}}
\newcommand{\Wsix}[6]{\left \{ \begin{array}{ccc} #1 & #2 & #3 \\ #4 & #5 & #6 \end{array}\right \} }
\newcommand{\Wnine}[9]{\left \{ \begin{array}{ccc} #1 & #2 & #3  \\ #4 & #5 & #6 \\ #7 & #8 & #9 \end{array}\right \} }

\usepackage[colorinlistoftodos,prependcaption,textsize=footnotesize]{todonotes}

\begin{document}

\title{Primordial fluctuations from quantum gravity: 16-cell topological model}

\author{Pietropaolo Frisoni}  
\email{pfrisoni@uwo.ca}
\affiliation{Dept.\,of Physics \& Astronomy, Western University, London, ON N6A\,3K7, Canada}

\author{Francesco Gozzini}
\email{gozzini@thphys.uni-heidelberg.de}
\affiliation{
Institut f\"ur Theoretische Physik, Philosophenweg 16, 69120 Heidelberg, Germany}

\author{Francesca Vidotto}
\email{fvidotto@uwo.ca}
\affiliation{Dept.\,of Physics \& Astronomy, Western University, London, ON N6A\,3K7, Canada}
\affiliation{Dept.\,of Philosophy and Rotman Institute, Western University, London, ON N6A\,3K7, Canada}

\begin{abstract}
\vskip1em
\noindent 
We present a numerical analysis of an Hartle-Hawking state for the early universe, in the deep quantum regime, computed using the covariant Loop Quantum Gravity formalism, in a truncation defined by 16-cell and in a simplified case where the dynamics is defined by $SU(2)$ BF theory. We compute mean geometry, fluctuations and correlations.  The results are consistent with the hypothesis that refining the triangulation does not affect the global physical picture substantially.
\end{abstract}

\maketitle

\section{Introduction}
\label{sec:introduction}

The early universe is believed to have emerged from a phase dominated by quantum gravitational effects.  A quantum theory of gravity is needed to describe this phase, and to  derive the initial boundary data of the cosmological standard  model.  An approach to address this derivation in the context of covariant Loop Quantum Gravity has been proposed in \cite{Gozzini_primordial}. It is based on two hypotheses.  First, the initial boundary data of the cosmological standard model can be computed as the result of a quantum gravitational transition amplitude from nothing as originally suggested by Hartle and Hawkings in the (different) context of Euclidean quantum gravity.  This state has been argued to be relevant both in the case of a genuine Big Bang and in the case of a Big Bounce   \cite{Lehners:2023yrj,Bianchi:2010zs,Vidotto:2011qa}. Second, a truncation of the theory to a finite number of degrees  of freedom capturing only a few lowest frequency modes at each cosmological time represents a good first order approximation in this regime.  These hypotheses allows us, in principle, to compute the initial state using the transition amplitudes of the theory. In particular, fluctuations and correlations of the resulting quantum state can be computed and compared with those of the quantum field theory initial state of standard cosmology. 

In practice, the calculation is hard. Initial numerical investigations have been developed in \cite{MCMC_paper}, relying on a drastic truncation, which can be interpreted as triangulation of the spacial geometry of a closed universe into five adjacent tetrahedra.  A serious limitation of this particular truncation is that all distinguishable regions of space are adjacent, hiding any dependence of correlations from spatial distance.

In this paper we take a step towards removing this limitation. We consider a finer truncation, corresponding to a triangulation of the geometry of a closed universe into sixteen tetrahedra.  In this triangulation the distinguishable regions of space are not all adjacent and we can analyse the dependence of the quantum correlations between different regions as a function on their separation.  

The paper focuses on the setting up of the combinatorics and on showing that the calculations are in principle doable numerical, but does not perform the calculations with the full covariant quantum gravity amplitudes, which will be explored in the future.   Here, instead, we use the unphysical simplified version of these amplitudes provided by the $SU(2)$ $BF$ model.  This is a rather drastic simplification of the amplitudes and therefore we present the results here more as a proof of concept than an actual test of the theory.  While there are regimes where the $BF$ amplitudes happen to be similar to the physical ones, this is not the case in general. 

In the next chapter we describe the relevant aspects of the geometry of a 16 cell.  Then we set of the calculation of the properties of the states that result from the transition to nothing. Finally, we give the numerical results showing the properties of this state.

\section{The 16-cell geometry}
\label{sec:16-cell_triangulation}

A 16-cell is the name of a regular polytope (the high-dimension generalization of a polyhedron) in four dimensional Euclidean space. It is the defined as the convex hull of the 8 points with euclidean coordinates $(0,0,0,\pm 1),(0,0,\pm 1,0),(0,\pm 1,0,0),(\pm 1,0,0,0)$. It is also called hexadecachoron and is analogous to the octahedron in three dimensions, namely the convex span of the points  $(0,0,\pm 1),(0,\pm 1,0),(\pm 1,0,0)$, and to the square in two dimensions, namely the convex span of the points  $(0,\pm 1),(\pm 1,0)$.  Its boundary is a regular triangulation of a 3-sphere into 16 tetrahedra.  The triangulation has 32 triangular faces, 24 edges, and the 8 vertices already mentioned. This is the second of the three regular triangulations of the 3-sphere: the first being the 4-simplex and the third being the 600-cell made by 600 tetrahedra.  The dual of this triangulation is the surface of an hypercube (which has 16 vertices). 

\begin{figure}[t]
    \centering
            \begin{subfigure}[b]{6cm}
       \includegraphics[width=5.3cm]{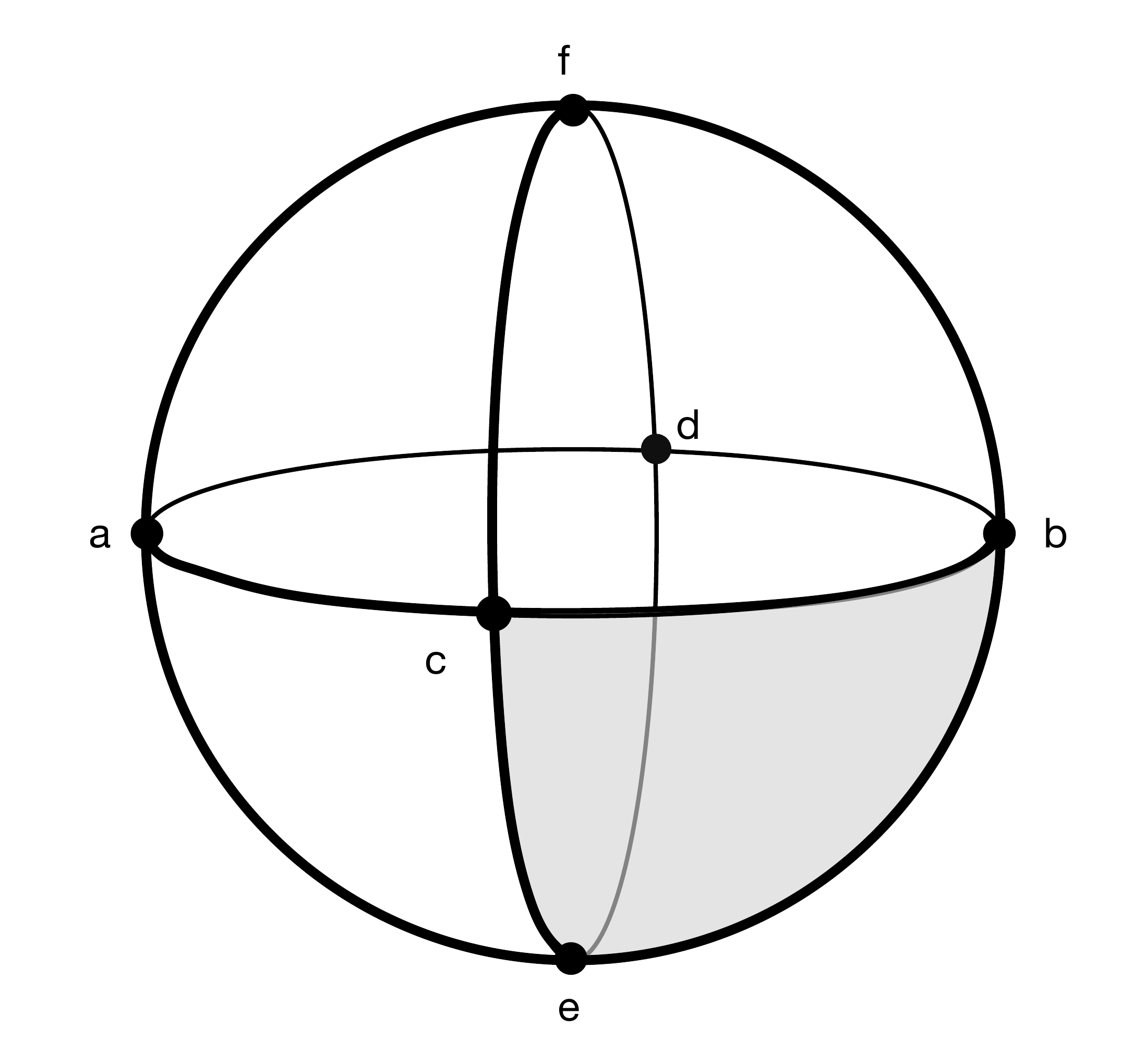}
     \end{subfigure}      
               \begin{subfigure}[b]{6cm}
     \includegraphics[width=5cm]{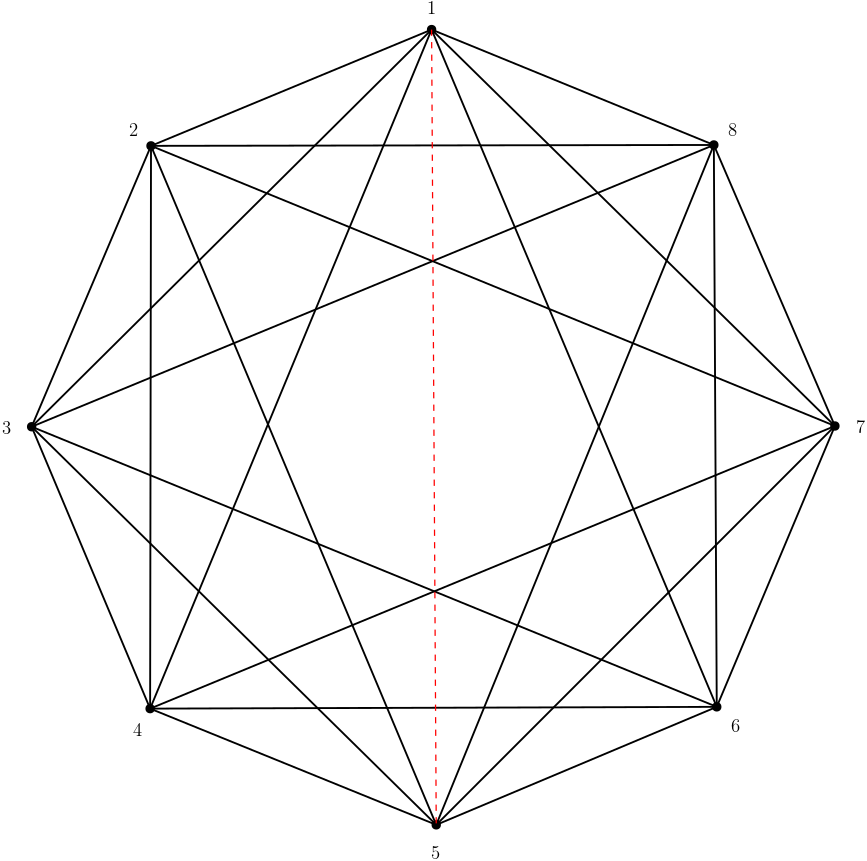}
     \end{subfigure}  
\caption{\label{fig:16cell} Triangulation of a 2-sphere with 8 triangles, one of which is highlighted in grey (left). Notice that this can be obtained by adding the points 5 and 6 to a triangulation of a 1-sphere (a circle) with four segments.  By repeating the same step one dimension higher, namely adding points  7 and 6 and connecting them to the 8 triangles, we obtain the 16-cell triangulation of a 3-sphere (right, with different numbering).}
\end{figure}

An intuitive visualization of the 16-cell triangulation is as follows.  Start with two points, say $a$ and $b$ (See Figure \ref{fig:16cell}).  By adding two extra points, say $c$ and $d$, each connected to both $a$ and $b$, we obtain a 1-sphere (a circle) split into 4 segments $S_1...S_4$, namely triangulated into a square.   By connecting each of these segments to two external points, say $e$ and $f$, we obtain 8 triangles that triangulate a 2-sphere into an octahedron, as in Fig. \ref{fig:16cell}.  By adding two more extra points, say $g$ and $h$ connecting each of them to the to these 8 triangles, we obtain 16 tetrahedra that triangulate a two sphere.  The two skeleton of the resulting triangulation is depicted in Figure \ref{fig:16cell} (with a different numbering of the vertices).

\medskip 


%
\begin{figure}[b]
    \centering
      \begin{subfigure}[b]{8.7cm}
     \includegraphics[width=6cm]{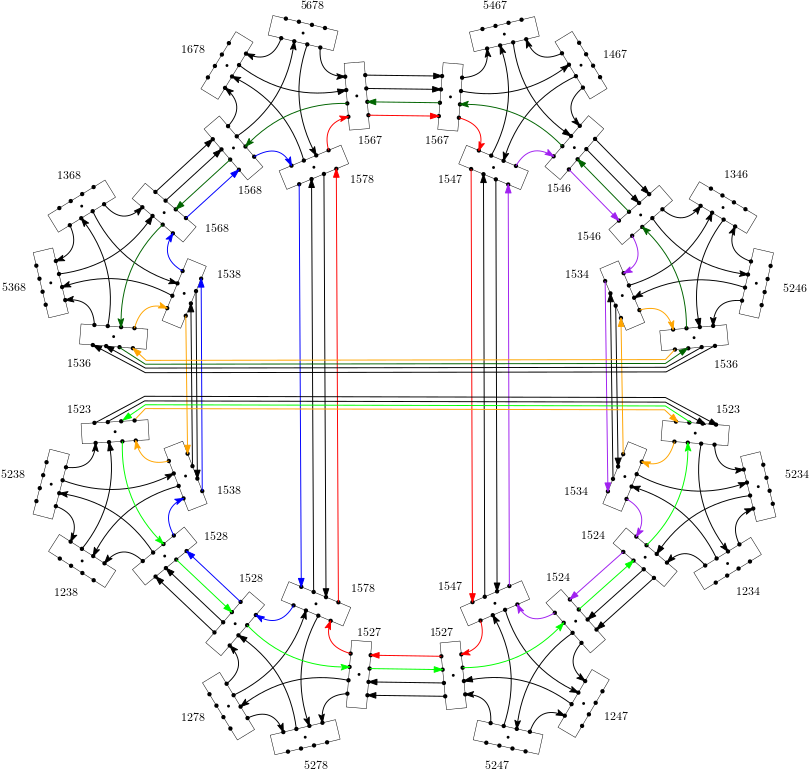}
     \end{subfigure}    
\caption{\label{fig:triangulation_and_spinfoam2}  {The spinfoam, with eight vertices and six internal faces (highlighted with different colors). The labels refer to the points in the triangulation. Only the edges are labelled not to clutter the picture.}}
\end{figure}

It is possible to triangulate the four dimensional 16-cell polytope by splitting it into eight 5-simplices.  Figure \ref{fig:triangulation_and_spinfoam2} illustrates this triangulation using the graphical notation of \cite{Alex}.  
This shows also how a spinfoam that defines an amplitude for this geometry can be computed \cite{libro}. 
The spinfoam has eight 4-simplices glued on six internal faces. The eight points in the triangulation are labeled using numbers $1, 2 \dots 8$. Consequently, in the  spinfoam diagram, each 4-simplex is labeled with five numbers, each tetrahedron with four numbers, and each face with three numbers. In Figure \ref{fig:triangulation_and_spinfoam2}, only the bulk and boundary tetrahedra  are labeled not to clutter the picture. As we discuss below, however, we shall not need the explicit form of this spinfoam. 

\section{Expectation values and correlations}
\label{sec:exp_values_and_corr}
\subsection{Boundary state}
\label{subsec:bound_state}
Let $\Gamma$ be an LQG graph with $L$ links and $N$ nodes. Then, the Hilbert space associated with $\Gamma$ is:
\begin{equation}
\label{eq:Hilbert_space_LQG}
\mathcal{H}_{\Gamma} = L_2 \left[ SU(2)^L / SU(2)^N \right] \ .
\end{equation}
The states $|j_l , i_{n} \rangle$ form the spin network basis in $\mathcal{H}_{\Gamma}$, where $n = 1 \dots N$,  $l = 1 \dots L$ (isolated indices range over multiple values). Half-integer spins constitute the set $j_l$, and $i_n$ is an intertwiner set. Each intertwiner $i$ is a basis element of the invariant subspace of the tensor product of $4$ $SU(2)$ representations at the corresponding node. 

Here we are interested in the Hilbert  $\mathcal{H}_{\Gamma}$ defined by the graph formed by the two skeleton of the triangulation of the boundary of the 16-cell. Hence $N=8$, $L=24$.  From now on, we omit the $\Gamma$ subscript not to weigh down the notation. 

Following \cite{Gozzini_primordial}, we fix all the spins to be equal so that $j_l = j$ for all the links of the boundary. Only one common spin is attached to all the links in this subspace. This define a subspace of $\mathcal{H}_{\Gamma}$ of the form
\begin{equation}
\label{HGamma}
H_\Gamma=\oplus_j \otimes_{n=1}^{16}  {\rm Inv}[V_j\otimes V_j\otimes V_j\otimes V_j],
\end{equation}
where $V_j$ is the spin-$j$ representation of $SU(2)$ and $\rm Inv$ denotes the $SU(2)$ invariant part of the tensor product. We write a boundary spin network state as: 
\begin{equation}
\label{eq:ia_def}
|j , i_n \rangle = |j, i_1 \rangle \otimes \dots \otimes |j, i_{16} \rangle  \ ,
\end{equation}
where $i_n$ are intertwiners in $ {\rm Inv}[V_j\otimes V_j\otimes V_j\otimes V_j]$.  Since all spins are equal to $j$, triangular inequalities constrain every intertwiner to assume only integer values between $0$ and $2j$. 

We are interested in the Hartle-Hawking state $| \psi_0 \rangle$ defined in \cite{Gozzini_primordial} and further studied in \cite{MCMC_paper}. In the Hilbert space \eqref{eq:Hilbert_space_LQG}, it is defined as:
\begin{equation}
\label{eq:cosm_state}
| \psi_0 \rangle = \sum_{j,i_n} \Wi  | j, i_n \rangle \ ,
\end{equation}
where $\Wi=\langle 0  | j, i_n \rangle_{Ph}$  is the LQG physical transition amplitude from the state defined by $j=0$.  This amplitude is associated with the transition from nothing to a boundary spin network state $| j, i_n \rangle$ with a given $j$.  The state \eqref{eq:cosm_state} can be interpreted as the natural state that is projected out of the vacuum state by the dynamics. 

In the $|j,i_n\rangle$ basis, the state reads  
\begin{equation}
\label{eq:state2}
\psi_0(j,i_n) =  \Wi.
\end{equation}
In quantum gravity, evolution is relative  with respect to one of the physical variables. It is natural in to use $j$ as the independent variable and interpret it as a proxy for a time variable, related to the overall size of the spacial 3-sphere. The $i_n$ are then interpreted as coding the local distortions of the geometry.  Then we can interpret $\psi_0(j,i_n)$ in analogy with the way we usually interpret $\psi(t,x)$: namely taking the first variable as labeling an evolution and the value of the state at fixed value of this variable as describing the quantum state at that given moment of the evolution.  In the following, we shall do so, thus studying 
\begin{equation}
\label{eq:state}
| \psi_0(j) \rangle = \sum_{i_n} \Wi  | j, i_n \rangle . 
\end{equation}
at different values of $j$. 

The sum in \eqref{eq:state} is over all the intertwiners in the set $i_n$, giving a total of $(2j + 1)^{16}$ elements entering the sum.  In Section \ref{subsec:computing_exp_values_MCMC}, we describe how to perform calculations with the state \eqref{eq:state} using Monte Carlo methods.
\subsection{Computing operators}
\label{subsec:computing_operators}
We define the expectation value on the state \eqref{eq:state} of a local operator $O$ over node $k$ as:
\begin{equation}
\label{eq:operator}
\langle O_k \rangle \equiv \frac{ \langle \psi_0 | O_k | \psi_0 \rangle }{\langle \psi_0 | \psi_0 \rangle} \ .
\end{equation}
The operator $O_k$ in \eqref{eq:operator} acts on the boundary $k$-th node of the Hilbert space \eqref{eq:Hilbert_space_LQG}. Using the orthogonality of the spin-network states 
\eqref{eq:ia_def}
\begin{equation}
\label{eq:iap_o_ia}
\langle j, i_n'| j, i_n \rangle   = \delta_{i_1', i_1} \ \dots \ \delta_{i_N', i_N} \ ,
\end{equation}
we define the normalization factor in the denominator of \eqref{eq:operator} as:
\begin{equation}
\label{eq:normalization_factor}
\langle \psi_0 | \psi_0 \rangle  = \sum_{i_n} \Wi^2 \equiv Z \ . 
\end{equation}
For a diagonal operator in the spin-network basis \eqref{eq:ia_def}, the expectation value \eqref{eq:operator} can be written as:
\begin{equation}
\label{eq:<On>}
\langle O_k \rangle = \Nf \sum_{i_n} \Wi^2 \langle j, i_k | O_k | j, i_k \rangle \ .
\end{equation}
In \eqref{eq:<On>}, the index $k$ is fixed  as it appears on both sides of the equations, while index $n$ ranges from $1$ to $N$. The expectation value of the product of two operators (on the node $k$ and $k'$ respectively) turns out to be:
\begin{equation}
\label{eq:<OnOm>}
\langle O_k O_{k'} \rangle = \Nf \sum_{i_n} \Wi^2 \langle j, i_k | O_k | j, i_k \rangle  \langle j, i_{k'} | O_{k'} | j, i_{k'} \rangle \ .
\end{equation}
We can now define the quantum spread of a local operator as:
\begin{equation}
\Delta O_k = \sqrt{\langle O_k^2 \rangle-\langle O_k \rangle^2} \ .
\label{eq:spread}
\end{equation}
Finally, we write the (normalized) correlation between two local operators as follows:
\begin{equation}
C(O_k, O_{k'}) = \frac{\langle O_k O_{k'} \rangle-\langle O_k \rangle \langle O_{k'} \rangle}{(\Delta O_k) \ (\Delta O_{k'})} \ .
\label{eq:correlations}
\end{equation}
\subsection{The 16-cell spinfoam amplitude}
\label{subsec:spinfoam_structure}

We now take a drastic simplification. Instead of the full covariant LQG amplitude $\Wi$
, we use the $SU(2)$ topological $BF$ amplitude
.  This simplifies the numerical calculation substantially and allows us to disregard a potential difficulty a problem: the six internal faces form a bubble in the spinfoam. (Bubbles have been recently studied numerically \cite{self_energy_paper, Dona_Frisoni_Ed_infrared, Frisoni_2023, VR_paper}.) 
Studying the 16-cell geometry using other non-topological spinfoam models (such as the EPRL theory) is left for future work. 

Thanks to the topological invariance of $BF$ theory, we can compute the amplitude by looking only at the boundary. This essentially reduces to computing a Wigner 48j-symbol (16 intertwiners plus 32 spins), because the amplitude $\Wi$ is simply the evaluation of the corresponding $SU(2)$ spin network. 

The steps for reducing the 48j symbol to simpler 21j symbols (each one decomposed in 6j and 9j Wigner symbols) are described below. This simplification allows us to avoid the introduction of an artificial cutoff on the spins labeling the internal faces in the bulk. The expression of the amplitude is greatly simplified, making it feasible to compute a huge amount of them.

We now describe the steps required to compute the 16-cell spinfoam amplitude. We start from the triangulation in the right panel of Figure \ref{fig:16cell}. We insert an internal segment between points 1 and 5 (arbitrarily chosen) to derive the corresponding dual triangulation, which is reported in Figure \ref{fig:triangulation_and_spinfoam2}. The spinfoam has eight vertices and six internal faces. Each vertex in the spinfoam is labeled with 5 points in the triangulation (each vertex is dual to a 4-simplex), each edge with 4 points (each edge is dual to a tetrahedron), and each face with 3 (each face is dual to a triangle). All tetrahedra in the triangulation labeled with points 1 and 5 are in the spinfoam bulk, while the remaining ones (labeled with points 1 or 5) are on the boundary. The tetrahedra in the triangulation sharing 3 points are glued on a face in the spinfoam. For example, the two tetrahedra labeled with points 5678 and 5278 are connected by a link on the boundary of the spinfoam.

\medskip

\begin{figure}[h]
    \centering
    \includegraphics[width=12cm]{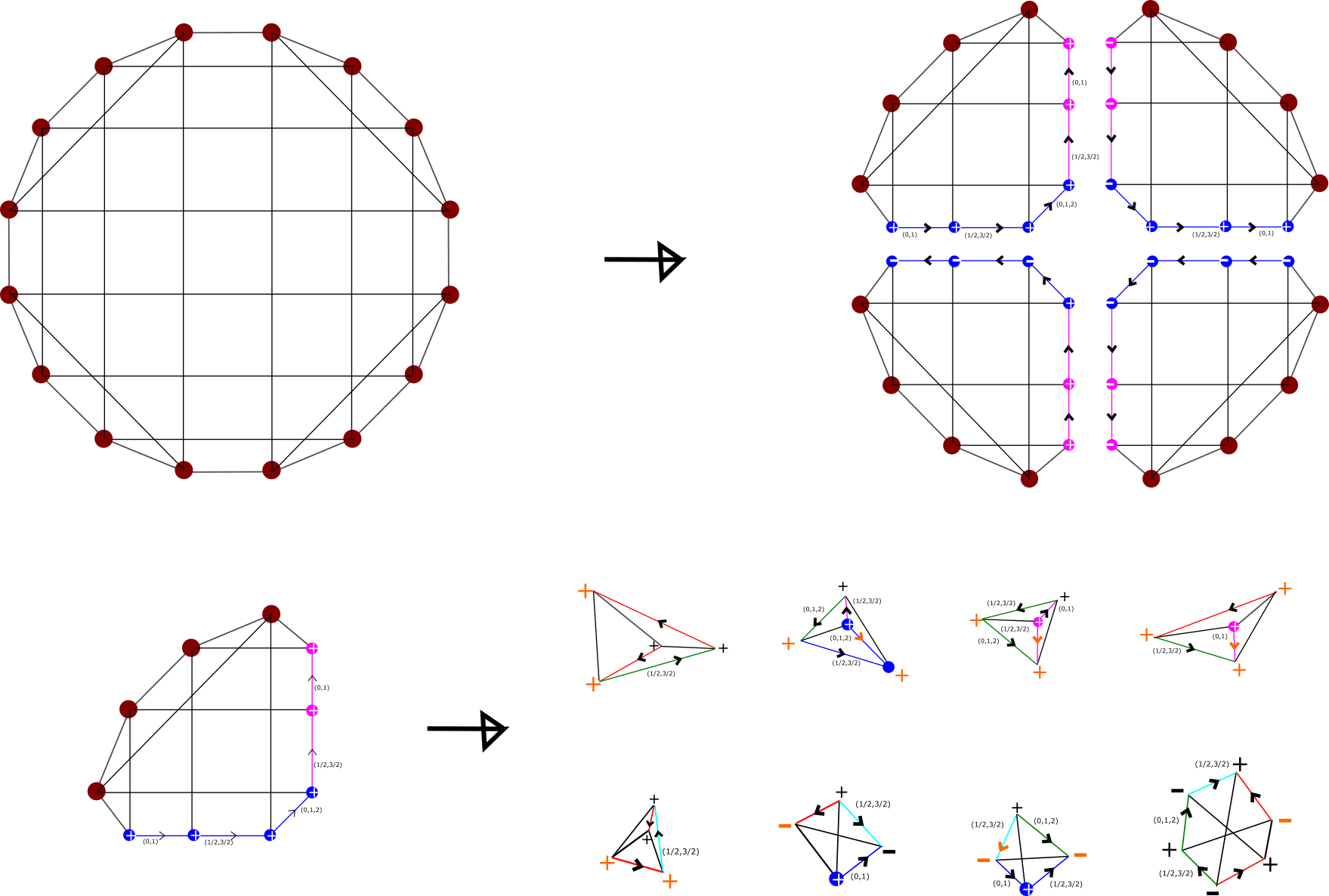}
    \caption{Top: \textit{The boundary of the 16-cell spinfoam amplitude is a 48j Wigner symbol, split as the contraction of four 21j symbols.} Bottom: \textit{Each 21j symbol is reduced to the sum of the products of seven 6j symbols and one 9j symbol.}} 
    \label{fig:16_cell_spinfoam_decomp}
\end{figure}

We first perform the integrals over $SU(2)$ in each one of the 16 boundary tetrahedra using relation \eqref{eq:from_box_to_intertwiners}, defined in \ref{app:SU(2)_coefficients}. Applying \eqref{eq:from_box_to_intertwiners} we have an intertwiner on each boundary tetrahedron of the spinfoam. Since the BF topological invariance allows computing the amplitude by focusing on the boundary, we perform the integrals over the $6$ internal faces of the 16-cell spinfoam. We are left with the diagram reported in the top left panel of Figure \ref{fig:16_cell_spinfoam_decomp}, where each boundary intertwiner is represented with a brown dot. 
A Wigner 48j symbol constitutes the boundary of the 16-cell spinfoam amplitude. To compute it efficiently, we write the symbol as contractions of smaller Wigner 21j symbols. The definition of the 21j symbol split as the contraction of 6j and 9j symbols is bulky. Therefore it is reported in \ref{app:SU(2)_coefficients} (see \eqref{eq:21j_exp}). We take advantage of it to write the ``north and south" amplitudes associated with the boundary of the top right panel in Figure \ref{fig:16_cell_spinfoam_decomp}. These are given by the contraction of 21j symbols along the vertical purple spins:
\begin{align}
\label{eq:W_N}
W_{N} & = \sum\limits_{p_1, p_2} \{ 21j \} \left( j, i_1, i_2, i_3, i_4, b_1, b_2, b_3, p_1, p_2 \right) \{ 21j \} \left( j, i_{16}, i_{15}, i_{14}, i_{13}, b_5, b_4, b_3, p_1, p_2 \right)  d_{p_1} d_{p_2}  \\
\label{eq:W_S}
W_{S} & = \sum\limits_{p_1, p_2} \{ 21j \} \left( j, i_8, i_7, i_6, i_5, b_1, b_2, b_3, p_1, p_2 \right) \{ 21j \} \left( j, i_9, i_{10}, i_{11}, i_{12}, b_5, b_4, b_3, p_1, p_2 \right) d_{p_1} d_{p_2} (-1)^{\chi} \ ,
\end{align}
where $\chi = 2 p_1 + 2 p_2 + 3 b_3$ and $d_{j_k} \equiv 2j_k + 1$. Finally, we contract the ``north and south" amplitudes \eqref{eq:W_N} - \eqref{eq:W_S} along the five horizontal blue spins in the top right panel of Figure \ref{fig:16_cell_spinfoam_decomp}. Therefore, we write the expression for the 16-cell BF spinfoam amplitude as:
\begin{align}
\label{eq:16_cell_amplitude}
W \left( j, i_n \right) & = \sum\limits_{b_1 \dots b_5} \left(  W_{N} \cdot  W_{S} \cdot d_{b_1} d_{b_2} d_{b_3} d_{b_4} d_{b_5} \right) \cdot \prod\limits_{k=1}^{16} \sqrt{ d_{i_k} } \ ,
\end{align}
where $i_n = i_1 \dots i_{16}$.
\section{Computational strategy}
\label{sec:computational_strategy}
We aim to compute the expectation values \eqref{eq:<On>}-\eqref{eq:<OnOm>}, the quantum spread \eqref{eq:spread}, and the correlation function \eqref{eq:correlations} for the 16-cell spinfoam described in section \ref{sec:16-cell_triangulation}. We want to do it for increasing values of $j$, which requires computing many 16-cell spinfoam amplitudes \eqref{eq:16_cell_amplitude}. Therefore, our first priority is efficiently computing the amplitude \eqref{eq:16_cell_amplitude} for each possible combination of boundary intertwiners. Next, we must assemble multiple amplitudes to calculate the quantities we are interested in.
\subsection{Computing the 16-cell spinfoam amplitude}
\label{subsec:computing_16-cell_amplitude}
As discussed in Section \ref{subsec:spinfoam_structure}, the topological BF 16-cell spinfoam amplitude can be written as a 48j Wigner symbol. Then, we can split the 48j Wigner symbol into a bunch of 21j symbols. Finally, each 21j symbol can be decomposed into 6j and 9j Wigner symbols. We proceed as follows to compute the amplitude numerically. 

\medskip 

\begin{figure}[h]
    \centering
   \includegraphics[width=12cm]{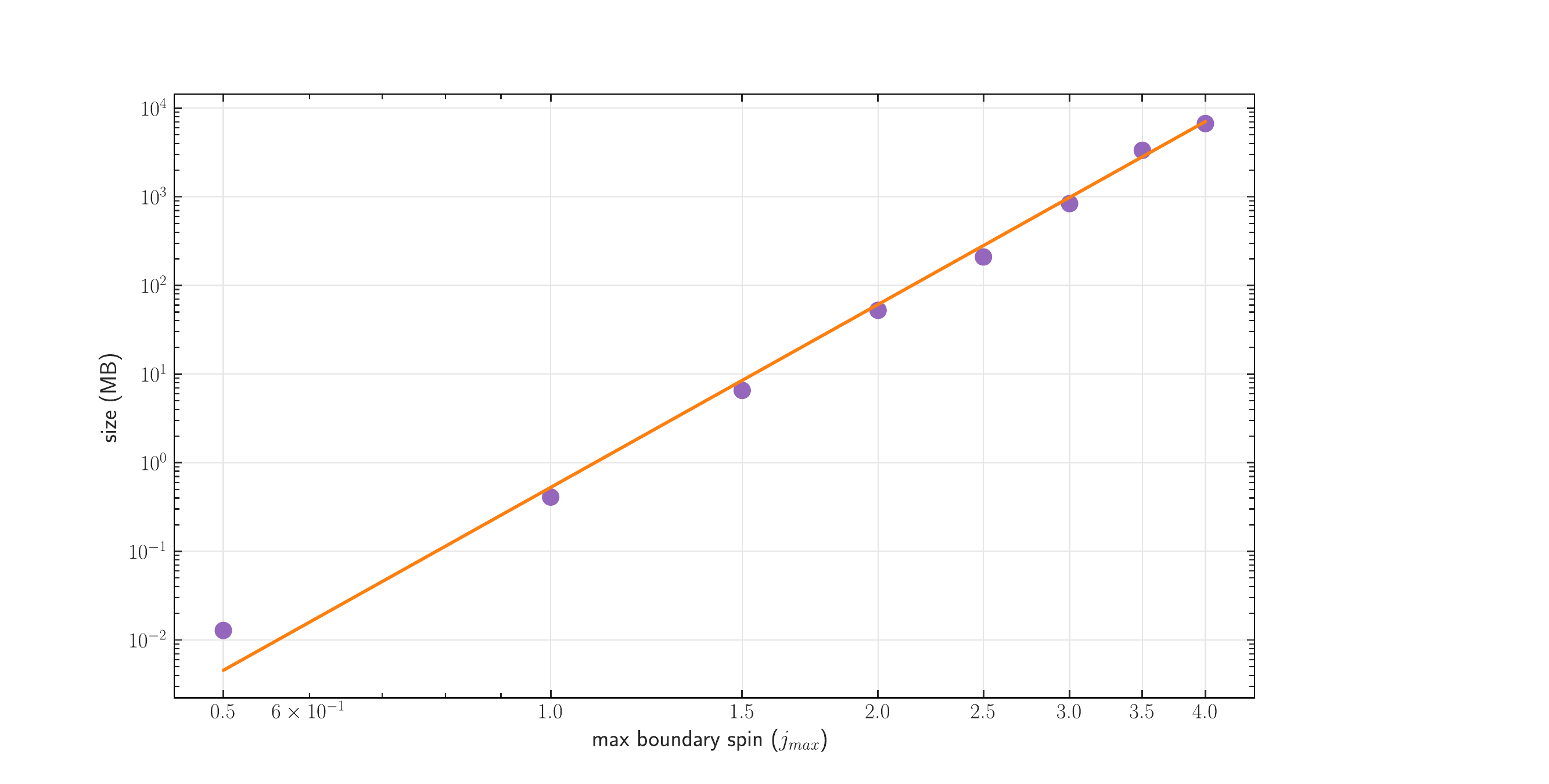}
     \caption{\label{fig:hash_table} \textit{Size of the computed parallel hash map tables of 21j Wigner symbols. Each symbol is stored in \textit{uint8} format to save memory. The hash table size roughly increases as $ j_{max}^{6.86}$.}}
\end{figure}

We first pre-compute all the 6j and 9j Wigner symbols with spins $j \leq 2j_{max}$. We do it for increasing values of $j_{max}$, relying on
\texttt{fastwigxj} library \cite{fastwigxj_related, Wigxjpf_library}. By default, it evaluates Wigner symbols quickly by lookup at precalculated tables, which are produced using the \texttt{wigxjpf} library. Then, we compute all the possible 21j symbols \eqref{eq:21j_exp} with spins $i_1, i_2, i_3, i_4, b_1, b_2, b_3, p_1, p_2$ less than or equal to a characteristic spin $\sim 2j_{max}$ (a combination of others bounds some of these, as triangular inequalities constrain the spins). Again, we do it for increasing $j_{max}$ values. The number of such symbols rapidly increases along with $j_{max}$. We store the 21j symbols using the \texttt{parallel hash map}, publicly available at \cite{parallel_hash_map}. The size of the hash tables as a function of $j_{max}$ is shown in Figure \ref{fig:hash_table}. 
By measuring the size of the calculated tables with a simple best polynomial fit, we conclude that the size roughly increases as:
\begin{equation}
\label{eq:size_hash_table}
\text{size} \approx j_{max}^{6.86} - 0.64 \ .
\end{equation}
Our strategy is to compute the 16-cell spinfoam amplitude \eqref{eq:16_cell_amplitude} by retrieving all the required 21j symbols from the hash tables. To do it efficiently, we must pre-load each table into the RAM or a cluster node-local storage before computing the amplitudes. The relation \eqref{eq:size_hash_table} imposes a major constraint on the required memory. Furthermore, as $j_{max}$ grows, the number of Wigner symbols to sum over and the time required to perform lookups in the hash table increase. Consequently, we must limit $j_{max}$ to very low values. We choose a maximum boundary spin $j_{max}=4$. This also corresponds to the deep quantum physical regime we want to investigate in this paper. 
\subsection{Computing expectation values with MCMC}
\label{subsec:computing_exp_values_MCMC}
We can now compute the 16-cell amplitude \eqref{eq:16_cell_amplitude} efficiently for different combinations of boundary intertwiners at fixed boundary spin $j$. We have the fundamental tool for computing the quantum expectation values and correlations. However, we need to address a major issue. In order to compute \eqref{eq:<On>} there are $(2j+1)^{16}$ summations to be performed. For $j=3$, there are $\sim 3.3 \cdot 10^{13}$ elements in the sum. Assuming (optimistically) that each amplitude \eqref{eq:16_cell_amplitude} can be computed in $\sim 10^{-5}$ seconds, we would need $\sim 10$ years to compute a single quantum expectation value! We can solve this issue using Markov Chain Monte Carlo (MCMC) over the boundary intertwiners. Specifically, we employ the Metropolis-Hastings algorithm \cite{MH_original_paper}. We report a brief introduction to the Metropolis-Hastings algorithm in \ref{app:M-H_review}, and we describe the details of the implementation below. This technique has been originally applied to spinfoams in \cite{MCMC_spinfoam_cosmology} in the case of a spinfoam model with $20$ boundary tetrahedra, to which we refer for further details. 

\medskip 

To compute \eqref{eq:<On>} using MCMC, we choose the following normalized target distribution over the state space \eqref{eq:ia_def}:
\begin{equation}
\label{eq:squared_prop_amp}
f_{j} \left( i_n \right) = \frac{ W^2 \left( j, i_{n} \right)}{\sum\limits_{i_n} W^2 \left( j, i_{n} \right)} \ .
\end{equation}
We use the Metropolis-Hastings algorithm to build a Markov chain with length $N_{mc}$ constituted by intertwiner states $[i_n]_1, [i_n]_2, \dots, [i_n]_{N_{mc}}$ so that each intertwiner state is generated from the distribution \eqref{eq:squared_prop_amp}. We use a truncated normal distribution rounded to integers as the proposal distribution, and each intertwiner (in every state) is sampled from a one-dimensional normal distribution. The center of the distribution is the value of the corresponding intertwiner in the previous state, truncated between $0$ and $2j$ (the range of all intertwiners). For example, if $j=2$ and $i_1 = 3$ in the state $[ i_n ]_s$, the proposed value for $i_1$ in the state $[ i_n ]_{s+1}$ is sampled from a normal distribution centered around $3$ and truncated between $0$ and $4$, and so on for the other intertwiners (each one is independent of others). We report in the flowchart \ref{numericalcode} below the full implementation of the algorithm to store Markov chains constituted by intertwiner states, using the target distribution \ref{eq:squared_prop_amp}. The parameters of the MCMC algorithm, the definition of the proposal distribution, and the expression of the truncated coefficients related to the proposal distributions are reported in the \ref{app:M-H_review}.
\begin{algorithm}[h]
\caption{Random walk over boundary intertwiners}\label{numericalcode}
\begin{algorithmic}[1]
\For{$j = 1 \dots j_{max}$}
\State Choose $N_{mc}$, the burn-in parameter $b$ and the standard deviation $\sigma$ as in \ref{tbl:MH_data}
\State Load into memory the hash table with the 21j Wigner symbols corresponding to $j$
\State Sample a random intertwiners configuration $[i_{n} ]_1$ and compute $W \left( j, [i_{n} ]_1 \right)$
\State Set initial multiplicity to $1$
\For{$s = 1 \dots N_{mc}$}
\State Generate a new state $[i_n ]$ from $[i_n ]_s$
\If{$[i_n ] = [i_n ]_s$} 
\State Increase the multiplicity by $1$
\State \textbf{continue}
\Else
\State Compute $W \left( j, [i_{n} ] \right)$ 
\State Compute $p = \textrm{min} \Bigl\{ 1 , \frac{W^2 ( j, [i_{n} ] )}{W^2 ( j, [i_{n} ]_s )} \frac{C_{0 , 2j, \sigma} ( [i_{n} ]_s)}{C_{0 , 2j, \sigma} ( [i_{n} ])} \Bigl\} $
\State Generate a uniform random number $r$ between $0$ and $1$
\If{$r < p$} 
   \If{$s > b$} 
      \State Store $[i_{n} ]_s$, $W \left( j, [i_{n} ]_s \right)$, and the corresponding multiplicity
   \EndIf
\State Set $[i_{n} ]_s \rightarrow [i_{n} ]$, $W \left( j, [i_{n} ]_s \right) \rightarrow W \left( j, [i_{n} ] \right)$  
\State Set the multiplicity to $1$
\Else
   \State Increase the multiplicity by $1$
\EndIf
\EndIf
\EndFor
\State Dump to disk all the states, amplitudes, and the corresponding multiplicities
\EndFor
\end{algorithmic}
\end{algorithm}

\begin{figure}[h]
    \centering
    \begin{subfigure}[b]{8.2cm}
    \includegraphics[width=8.5cm]{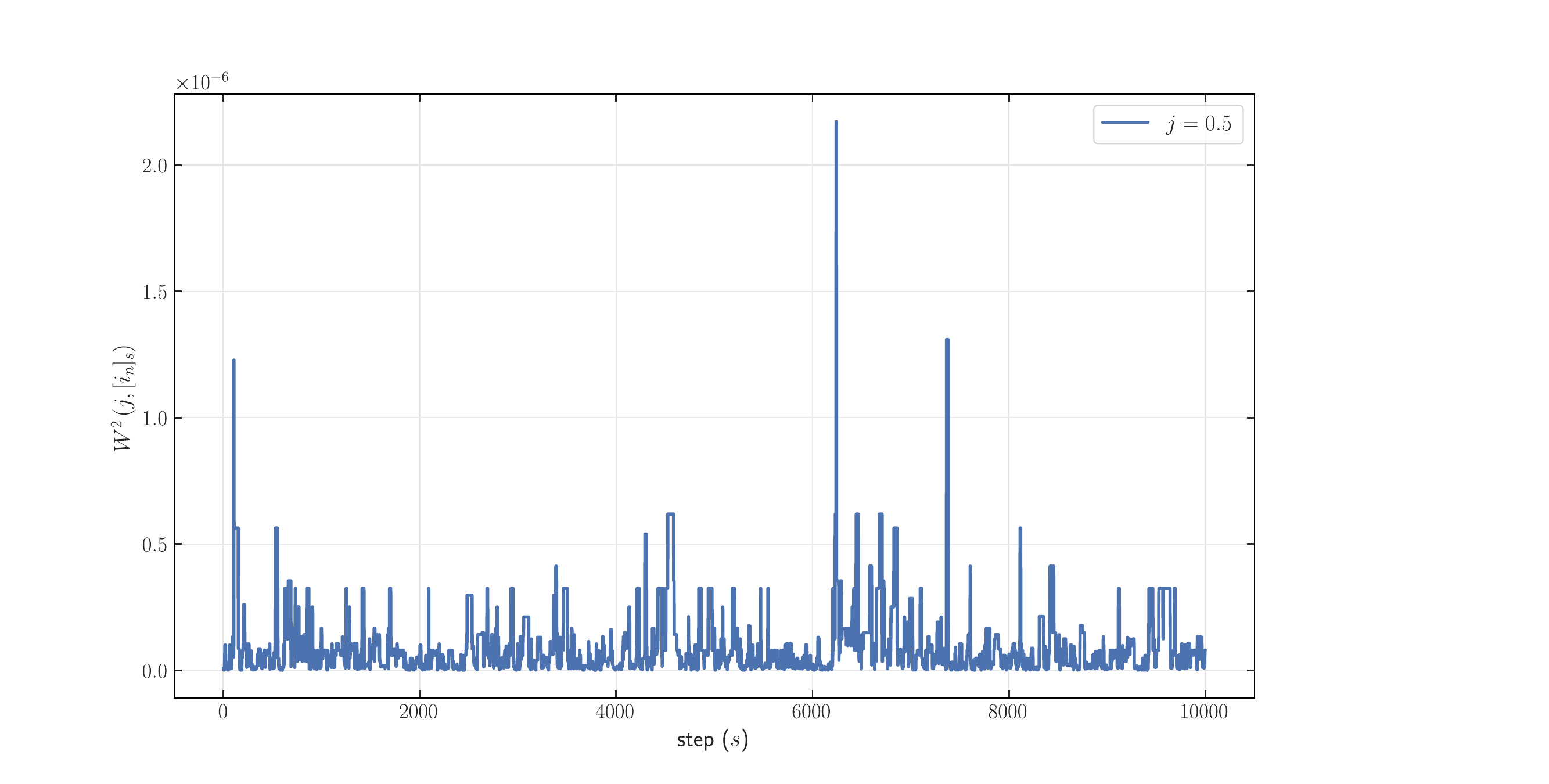}
     \end{subfigure}
     \begin{subfigure}[b]{8.2cm}
     \includegraphics[width=8.5cm]{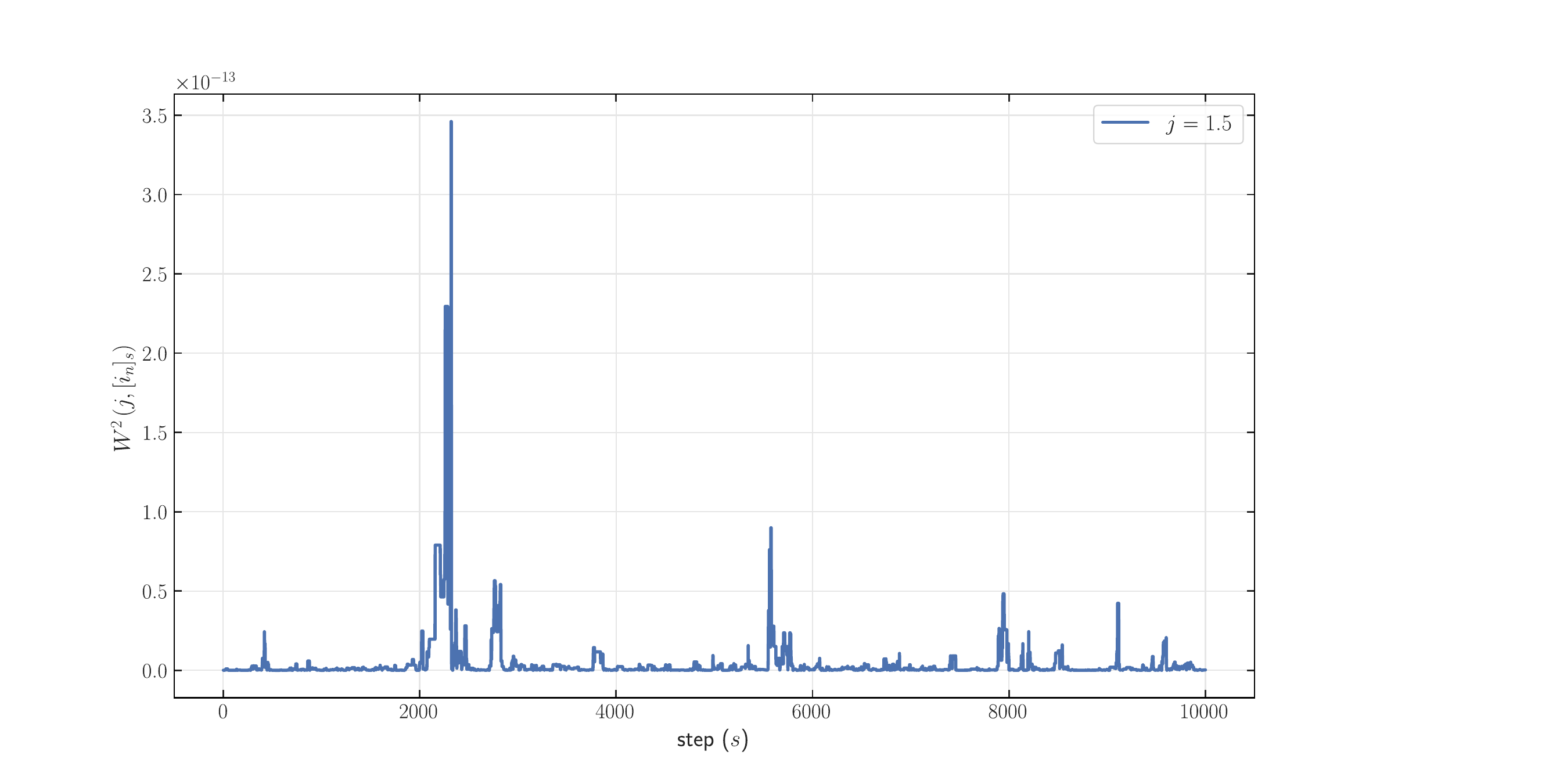}
      \end{subfigure}
     \begin{subfigure}[b]{8.2cm}
    \includegraphics[width=8.5cm]{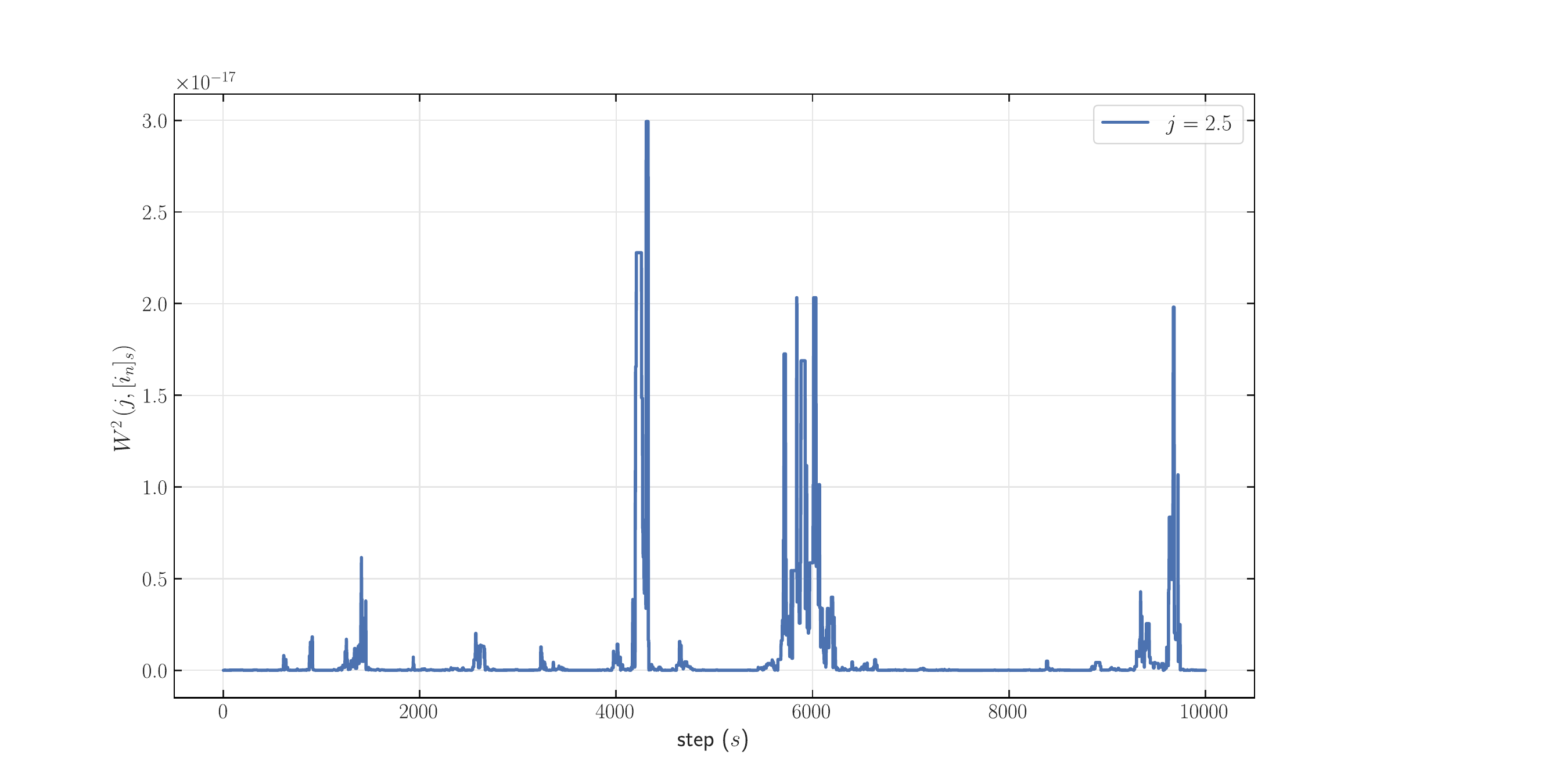}
      \end{subfigure}
       \begin{subfigure}[b]{8.2cm}
    \includegraphics[width=8.5cm]{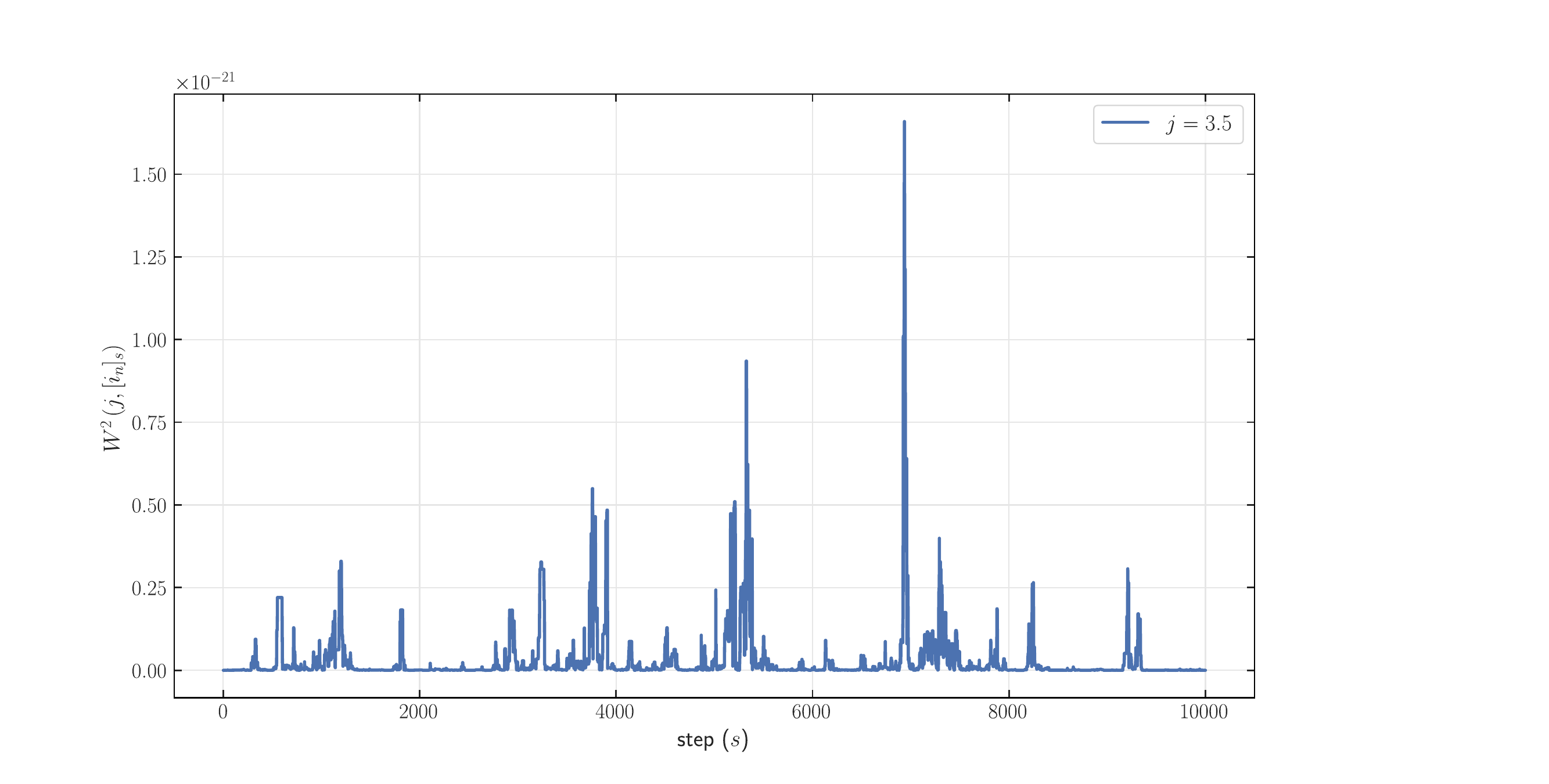}
      \end{subfigure}    
    \caption{\label{fig:amplitude_MCMC} \textit{Squared amplitude $W^2 \left( j, [i_{n} ]_s \right)$ computed in the algorithm \ref{numericalcode} as function of the number of steps $s$ along the Markov Chain. The frequency of amplitude's peaks tends to decrease as j increases.}}   
\end{figure}

We show in Figure \ref{fig:amplitude_MCMC} the squared amplitude $W^2 \left( j, [i_{n} ]_s \right)$ as a function of the number of steps $s$ for the first $10^4$ iterations of the MCMC algorithm \ref{numericalcode}. We notice that the ``hotspots" become less and less frequent as $j$ increases. This is most probably because the space of states becomes larger so that the sampler requires more iterations to find the relevant contributions. To overcome this issue, an interesting approach (currently under development) consists in replacing MCMC in spinfoams with GFlowNets \cite{GFL_Net_paper, GFL_Net_paper_2}. We tried many different values for the burn-in parameter $b$ (ranging from $0$ to $10^4$) and found no significant dependence. This can be because the distribution \eqref{eq:squared_prop_amp} has many isolated peaks regularly distributed in the space \eqref{eq:cosm_state}. The time complexity of algorithm \ref{numericalcode} depends on the proposal distribution's standard deviation $\sigma$. In fact, the higher $\sigma$ is, the higher the probability that the sampler moves from the current state is high (therefore, a new amplitude must be computed). Moreover, the standard deviation of the proposal distribution affects the acceptance rate: the lower $\sigma$ is, the higher the acceptance rate becomes (in the extreme case $\sigma = 0$, the sampler never moves, and all proposed states are accepted). We balance $\sigma$ so that the acceptance rate of the Metropolis-Hastings algorithm ranges between $\sim 30 \%$ and $\sim 45 \%$. 

\medskip 

Next, we use the intertwiner draws stored during the algorithm \ref{numericalcode} to evaluate the expectation value \eqref{eq:<On>} by applying the Monte Carlo summation \cite{VR_paper}:
\begin{equation}
\label{eq:<On>_MC}
\langle O_k \rangle_{mc} = \frac{1}{ N_{MC}} \sum\limits_{s=1}^{N_{mc}} \langle j , [i_k]_s | O_k | j, [i_k]_s \rangle \ . 
\end{equation}
With respect to \eqref{eq:<On>}, we replaced a sum over $(2j+1)^{16}$ intertwiners with a sum over $N_{mc}$ elements. This hugely simplifies the calculation, which would be impractical otherwise. The application of Monte Carlo to the corresponding quantum spread \eqref{eq:spread} and to the correlation functions \eqref{eq:correlations} is straightforward:
\begin{equation}
\label{eq:spread_MC}
\Delta \langle O_{k} \rangle_{mc} =  \sqrt{\langle O_k^2 \rangle_{mc}-\langle O_k \rangle_{mc}^2} \ ,   
\end{equation}
\begin{equation}
 C( O_{k} , O_{k'})_{mc} = \frac{\langle O_k O_{k'} \rangle_{mc}-\langle O_k \rangle_{mc} \langle O_{k'} \rangle_{mc}}{(\Delta \langle O_{k} \rangle_{mc}) \ (\Delta \langle O_{k'} \rangle_{mc})} \ .   
\end{equation}
We want to estimate the error in \eqref{eq:<On>_MC} due to the statistical fluctuations of MCMC. For this purpose, we repeat the calculation of \eqref{eq:<On>} multiple times using different (and independent) Markov chains with the same parameters. Then, we compare the results, allowing us to improve the estimation and quantify the fluctuations. If we have $C$ different estimates $\langle O_k \rangle_{mc}^{(1)}, \ \langle O_k \rangle_{mc}^{(2)}, \dots \, \langle O_k \rangle_{mc}^{(C)}$, we compute the corresponding average:
\begin{equation}
\label{eq:average_<O>_MC}
\overline{\langle O_k \rangle}_{mc} = \frac{1}{C} \sum\limits_{c=1}^{C} \langle O_k \rangle_{mc}^{(c)} \ .
\end{equation}
For completeness, we write the corresponding average for the spread and correlations explicitly:
\begin{equation}
\label{eq:spread_MC_average}
\overline{\Delta \langle O_{k} \rangle}_{mc} =  \frac{1}{C}\sum\limits_{c=1}^{C} \Delta \langle O_{k} \rangle_{mc}^{(c)} \ ,   
\end{equation}
\begin{equation}
\label{correlation_MC_average}
\overline{C( O_{k} , O_{k'} )}_{mc} = \frac{1}{C} \sum\limits_{c = 1}^{C} C( O_{k} , O_{k'} )_{mc}^{(c)} \ . 
\end{equation}
To have an intuitive visualization, we consider the normal distribution associated with the average \eqref{eq:average_<O>_MC}. For this purpose, we introduce the standard deviation between the expectation values over the chains:
\begin{equation}
\label{eq:sigma}
\sigma = \sqrt{\frac{1}{C}\sum_{c=1}^{C} \left( \overline{\langle O_k \rangle}_{mc} - \langle O_k \rangle_{mc}^{(c)}  \right)^2} \ ,
\end{equation}
so that we can write the normal distribution as:
\begin{equation}
\label{eq:gaussian}
\mathcal{N}_{O} (x) = \frac{1}{\sigma \sqrt{2\pi}} e^{- \frac{1}{2} \left( \frac{x - \overline{\langle O_k \rangle}_{mc}}{2 \sigma } \right)^2} \ .
\end{equation}
If the standard deviation \eqref{eq:sigma} is low, the statistical fluctuations in the estimate \eqref{eq:average_<O>_MC} are reasonably under control. We notice that increasing $C$ does not reduce \eqref{eq:sigma}, as the standard deviation is reduced by increasing the number of samples $N_{mc}$ along each chain. 

\medskip 

Therefore, we must run the algorithm \ref{numericalcode} $C$ times to store as many independent Markov Chains. Since each chain is independent, we can parallelize the code across multiple threads using the C/C++ OpenMP library \cite{chandra2001parallel}. Each thread runs a different Markov chain with the given parameters, storing the computed intertwiner draws. The parallel hash table with all the required 21j Wigner symbols is loaded only once into the RAM so that each thread retrieves the symbols \eqref{eq:21j_exp} from the same table. By doing this, we can compute the average over $C$ expectation values \eqref{eq:average_<O>_MC}, spread \eqref{eq:spread_MC_average} and correlations \eqref{correlation_MC_average} computed with as many independent chains. For this task, we take advantage of the Python libraries \texttt{Pandas} and \texttt{Numpy} \cite{harris2020array, mckinney2010data}. As reported in table \ref{tbl:MH_data}, in this paper, we computed and stored $C=30$ Markov chains, each with $10^6$ intertwiner draws.    

\medskip 

Finally, we notice that the expression of the 16-cell spinfoam amplitude \eqref{eq:16_cell_amplitude} requires a considerable numerical effort to be computed, despite the strategy described in Section \ref{subsec:computing_16-cell_amplitude}. This limitation allows us to compute only diagonal boundary operators. For this class of operators, the only amplitudes \eqref{eq:16_cell_amplitude} we need to compute are those required in the algorithm \ref{numericalcode}. In the case of non-diagonal operators, such as the volume of boundary tetrahedra, we would need to compute $2j$ additional amplitudes \eqref{eq:16_cell_amplitude} for each term in the sum \eqref{eq:<On>_MC}.
\section{Numerical results}
\label{sec:results}
In this Section, we report the numerical values obtained for the expectation values \eqref{eq:<On>}, the quantum spread \eqref{eq:spread}, and correlations \eqref{eq:correlations} following the procedure described in Section \ref{subsec:computing_exp_values_MCMC}. Here we focus on the dihedral angle operator, describing the external dihedral angle between faces $a$ and $b$ of each boundary tetrahedron. Its expression in the basis states \eqref{eq:cosm_state} is \cite{Gozzini_primordial}:
\begin{equation}
  \label{eq:geom-angleformula}
 \langle j, i_n  | \cos(\theta)_k | j, i_n \rangle = \frac{i_k(i_k+1) - 2j(j+1) }{2 j(j+1)} \ .
\end{equation}
Due to its simplicity and geometrical interpretation, the dihedral angle operator \ref{eq:geom-angleformula} is an optimal candidate to investigate the geometrical properties of the 16-cell triangulation described in Section \ref{sec:16-cell_triangulation}. Furthermore, it is the same operator studied in \cite{Gozzini_primordial, MCMC_paper}, which allows immediate comparison with previously studied triangulations.
\subsection{Expectation values}
\label{subsec:exp_values}
\begin{figure}[H]
    \centering
    \begin{subfigure}[b]{8.2cm}
    \includegraphics[width=8.5cm]{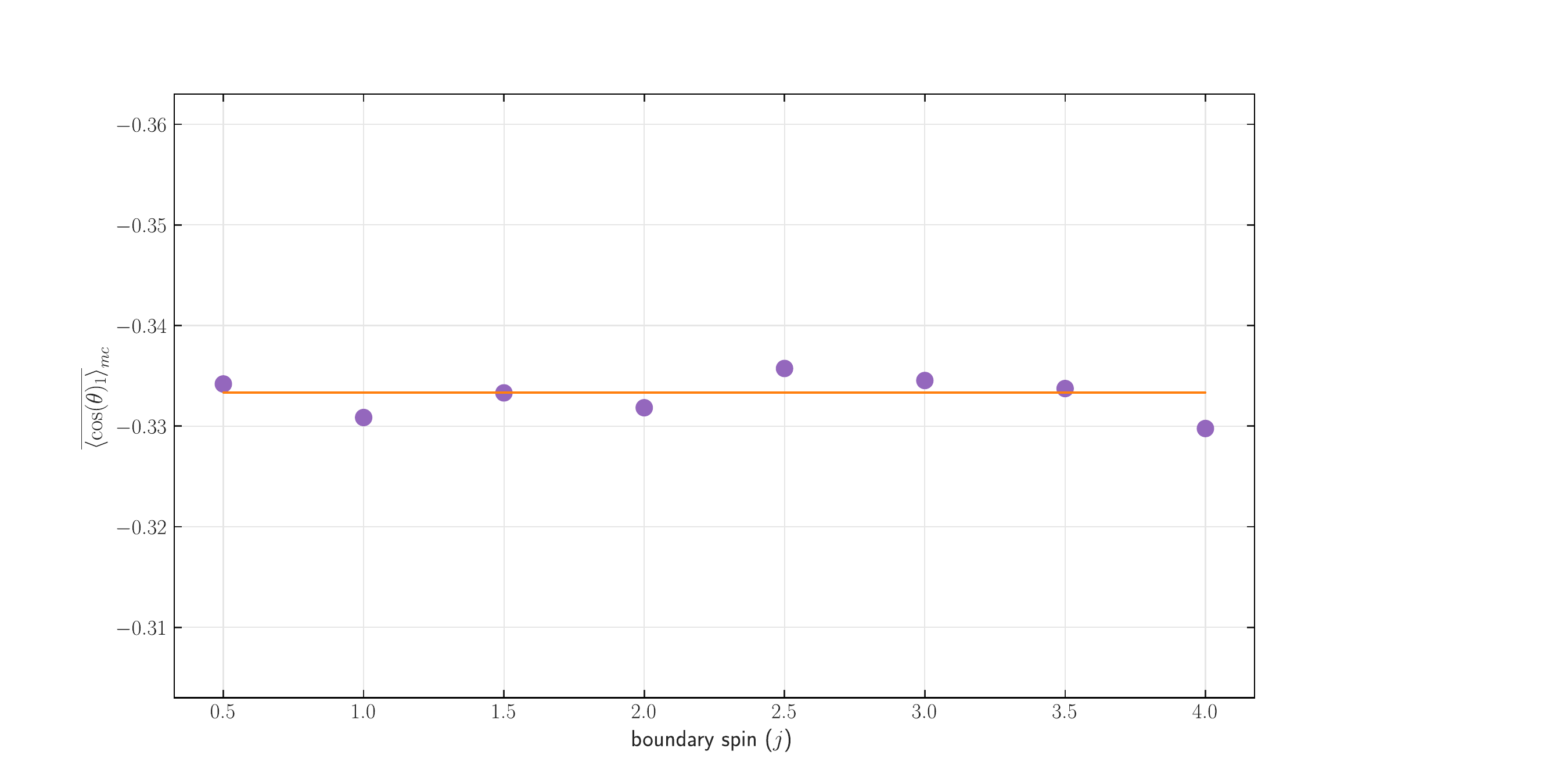}
     \end{subfigure}
     \begin{subfigure}[b]{8.2cm}
     \includegraphics[width=8.5cm]{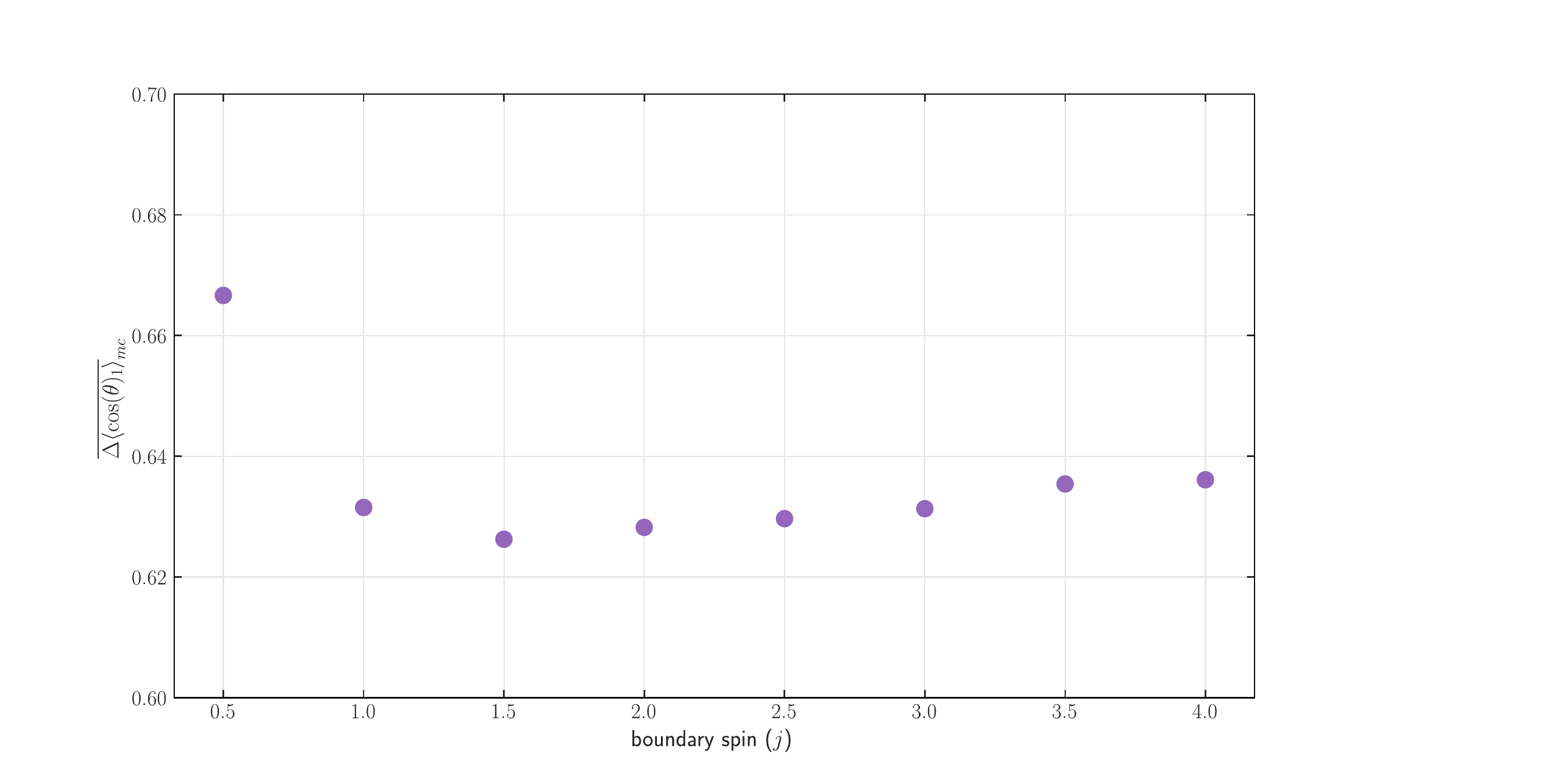}
      \end{subfigure}
    \caption{\label{fig:angles}Left: \textit{expectation values \eqref{eq:average_<O>_MC} of the dihedral angle operator \eqref{eq:geom-angleformula}. We show only the result for the first node, but all other nodes have the same behavior. The orange line shows the value of the cosine of the external dihedral angle of a regular tetrahedron $\arccos(\frac{1}{3})$.} Right: \textit{corresponding quantum spread \eqref{eq:spread_MC_average}.}}   
\end{figure}

The expectation value \eqref{eq:average_<O>_MC} of the dihedral angle operator \eqref{eq:geom-angleformula} and the quantum spread \eqref{eq:spread_MC_average} are shown in Figure \ref{fig:angles}. Both these quantities have been computed averaging over $C=30$ independent Markov chains. 
A few comments are in order. The first observation is that the expectation value \eqref{eq:geom-angleformula} of all 16 boundary tetrahedra is peaked on the value of an external dihedral angle of an equilateral tetrahedron. This coincides with the result obtained with the simplest possible triangulation of a 3-sphere \cite{Gozzini_primordial} and the triangulation used in \cite{MCMC_paper}. Unlike the 4-simplex studied in \cite{Gozzini_primordial}, the 16-cell is not self-dual. This dynamical result of the global geometry confirms that the metric of the state \eqref{eq:cosm_state} averages to that of a regular 3-sphere even in the 16-cell geometry and is not a consequence of the reduction \eqref{eq:ia_def}. The second observation is that the quantum spread is slightly increasing as a function of the boundary spin $j$, which indicates that the quantum fluctuations are ample. This agrees with the result obtained in \cite{MCMC_paper} in the topological model. It is reasonable to expect larger fluctuations for the EPRL model. We leave the analysis of the EPRL case for future work. 

\medskip

\begin{figure}[H]
    \centering
      \includegraphics[width=12cm]{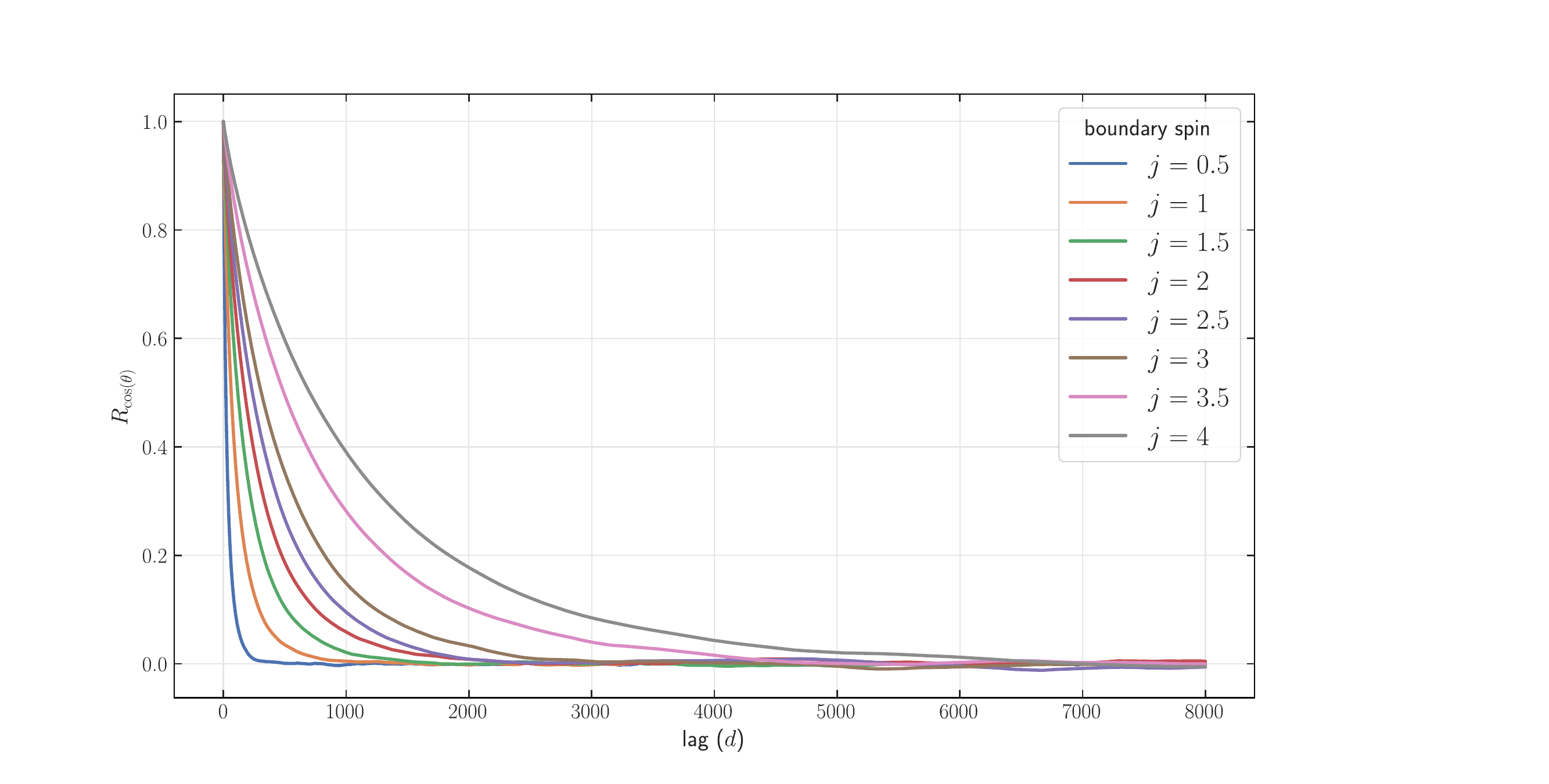}
    \caption{{\label{fig:angles_autocorrelation} \textit{Autocorrelation function \eqref{eq:autocorrelation} of the expectation value \eqref{eq:<On>_MC} of the dihedral angle operator \eqref{eq:geom-angleformula}. The autocorrelation decays more slowly as a function of the lag $d$ as the boundary spin $j$ increases.}}}
    \end{figure}

As described in \ref{app:M-H_review}, a non-zero correlation exists between the intertwiner states stored during the algorithm \ref{numericalcode}. This is a consequence of the Markovian nature of the process. We want to check whether the quality and number of states generated with the algorithm \eqref{numericalcode} are sufficient to accurately approximate the target distribution \eqref{eq:squared_prop_amp}. For this purpose, we compute the autocorrelation function \eqref{eq:autocorrelation} for the matrix elements of the dihedral angle operator \eqref{eq:geom-angleformula} using the states generated with the Metropolis-Hastings algorithm. These terms define the expectation values \eqref{eq:<On>_MC}. We computed the autocorrelation function over all the 30 chains stored and for all nodes, finding very similar behavior. We report explicitly in Figure \ref{fig:angles_autocorrelation} the autocorrelation function computed using chain one and considering the first node of the 16-cell spinfoam. The result clearly shows that the autocorrelation is large at short lags but goes to zero pretty fast (we show a maximum lag of $8000$ for visualization purposes, but each chain contains $10^6$ states). This is precisely what we expect for a Markov chain that converges to a stationary distribution, indicating that the sample size that we considered is sufficient. Interestingly, the autocorrelation's decaying becomes slower as the boundary spin $j$ increases. 

\medskip

\begin{figure}[h]
    \centering
    \includegraphics[width=12cm]{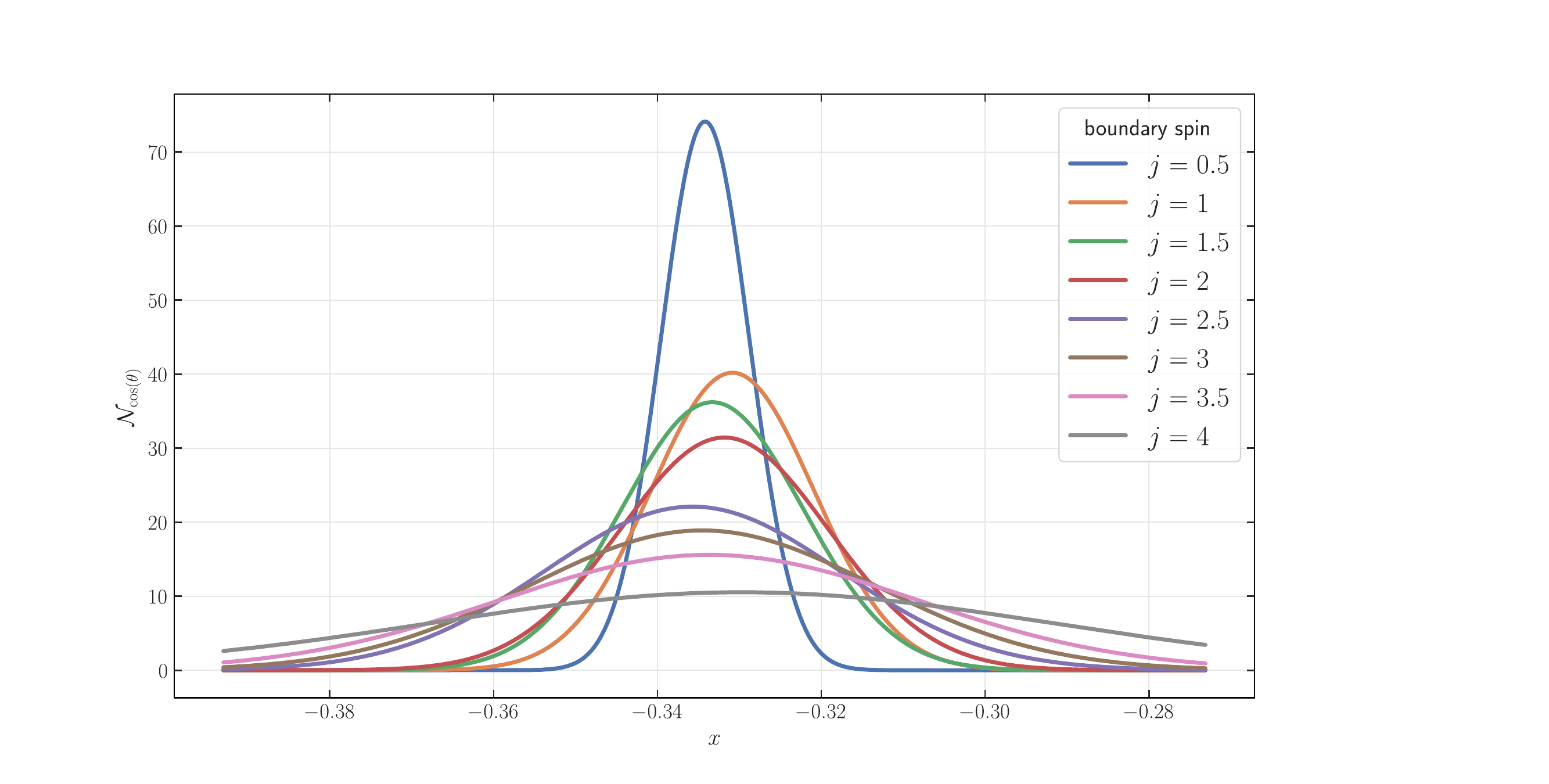}
    \caption{{\label{fig:angesl_numerical_fluctuations} \textit{Normal distribution \eqref{eq:gaussian} of the expectation values of the dihedral angle operator \eqref{eq:geom-angleformula}. The statistical fluctuations increase as a function of the boundary spin $j$.}}}
\end{figure}

As a final check for the effectiveness and reliability of the MCMC algorithm described in Section \ref{subsec:computing_exp_values_MCMC}, we repeat the calculation of the dihedral angle operator multiple times using equation \eqref{eq:average_<O>_MC}. The plot of the corresponding normal distribution \eqref{eq:gaussian} is reported in Figure \ref{fig:angesl_numerical_fluctuations}. For convenience, we report the plot considering node 1 of the 16-cell spinfoam, but we performed the calculation over all 16 nodes finding a very similar behavior. Interestingly, the statistical fluctuations become larger as the boundary spins $j$ increase. This confirms that we have enough control over the stochastic fluctuations due to Monte Carlo.

\newpage
\subsection{Correlations}
\label{subsec:correlations}
\begin{figure}[H]
   \begin{subfigure}[b]{6cm}
        \includegraphics[width=5.6cm]{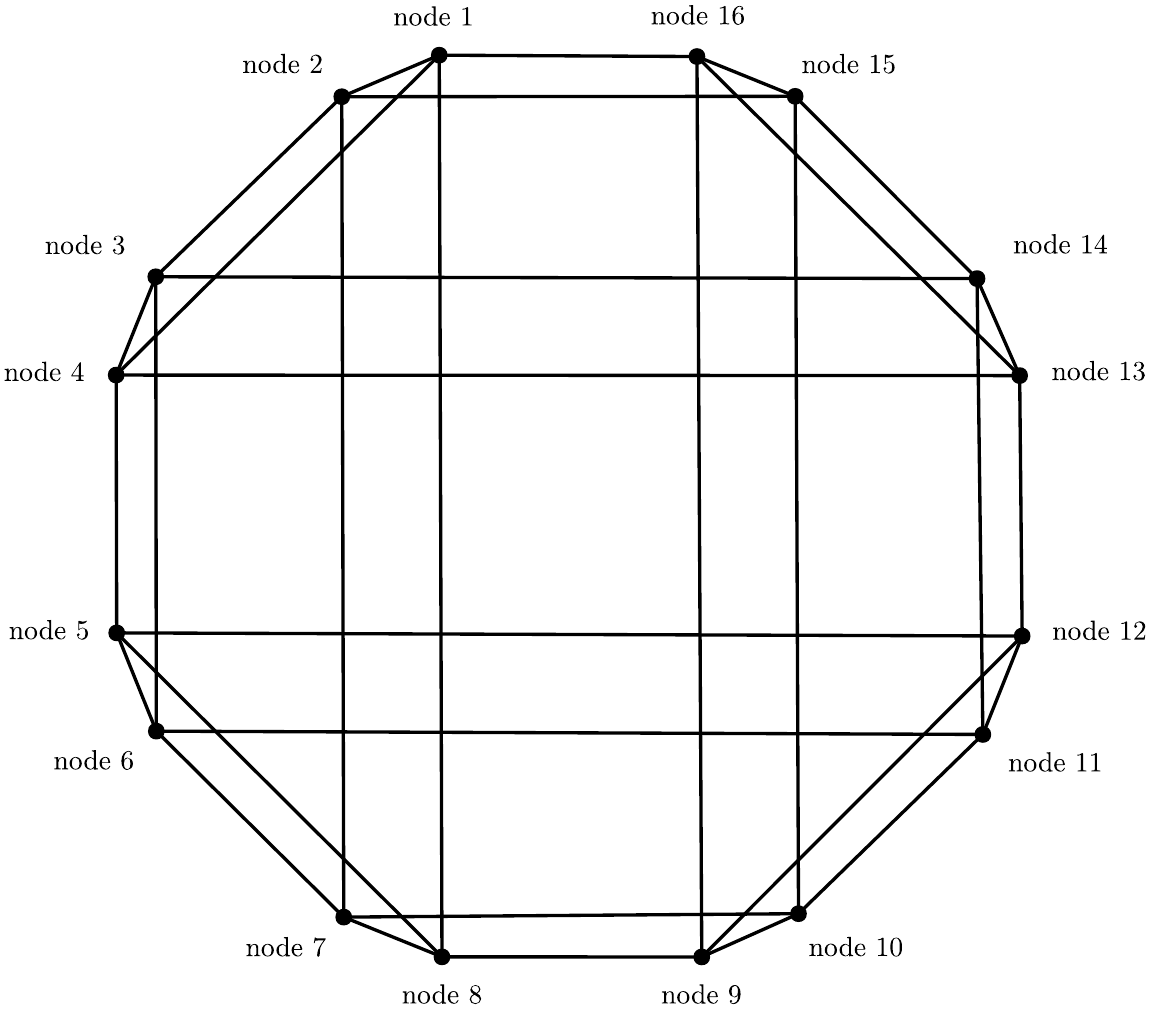}
        \end{subfigure}  
    \begin{subfigure}[b]{9cm}
        \includegraphics[width=11cm]{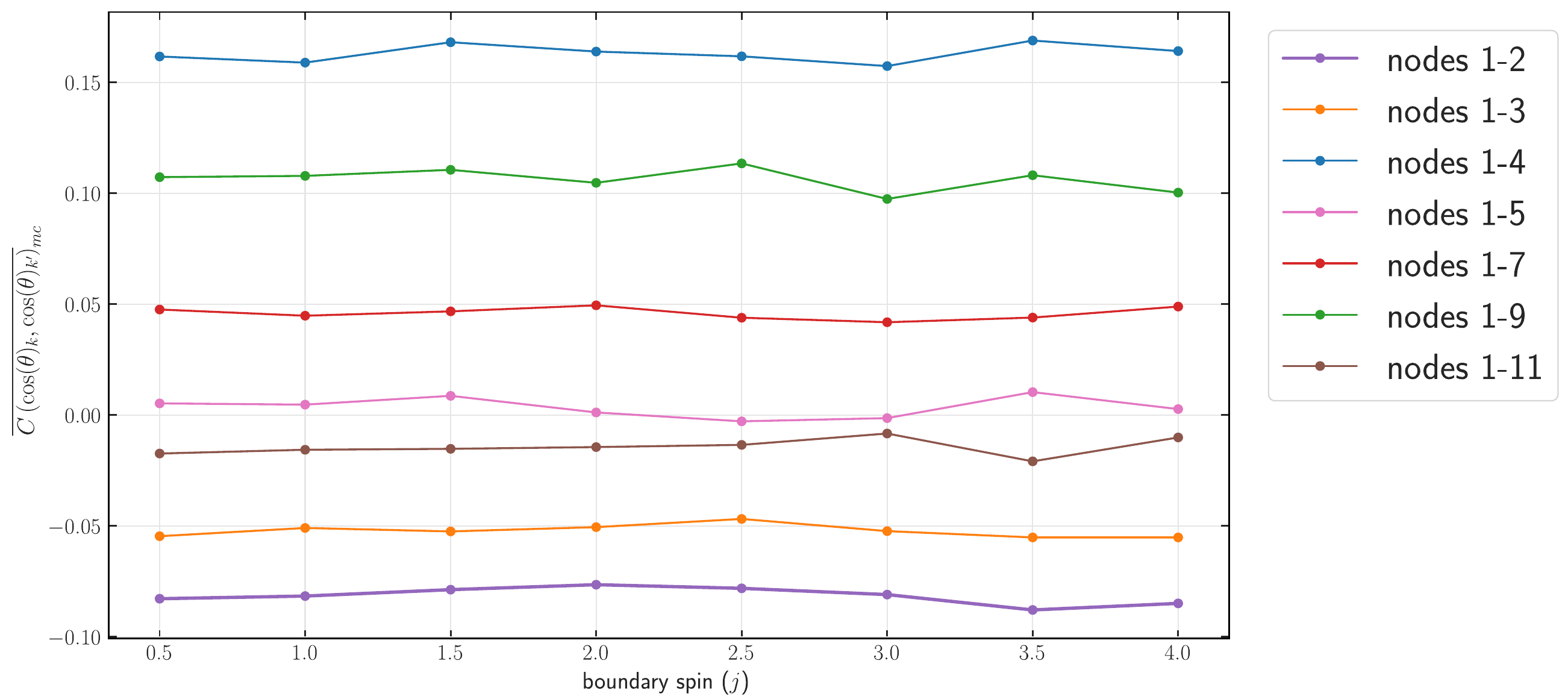}
        \end{subfigure}       
    \caption{{\label{fig:angles_correlations} Left: \textit{16-cell spinfoam boundary (same as Figure \ref{fig:16_cell_spinfoam_decomp}) with nodes labeled by numbers} Right: \textit{Some values of quantum correlations \eqref{correlation_MC_average} as a function of the boundary spin $j$.}}}
    \end{figure}

\begin{figure}[H]
    \centering
    \begin{subfigure}[b]{8.2cm} \!\!
    \includegraphics[width=10cm]{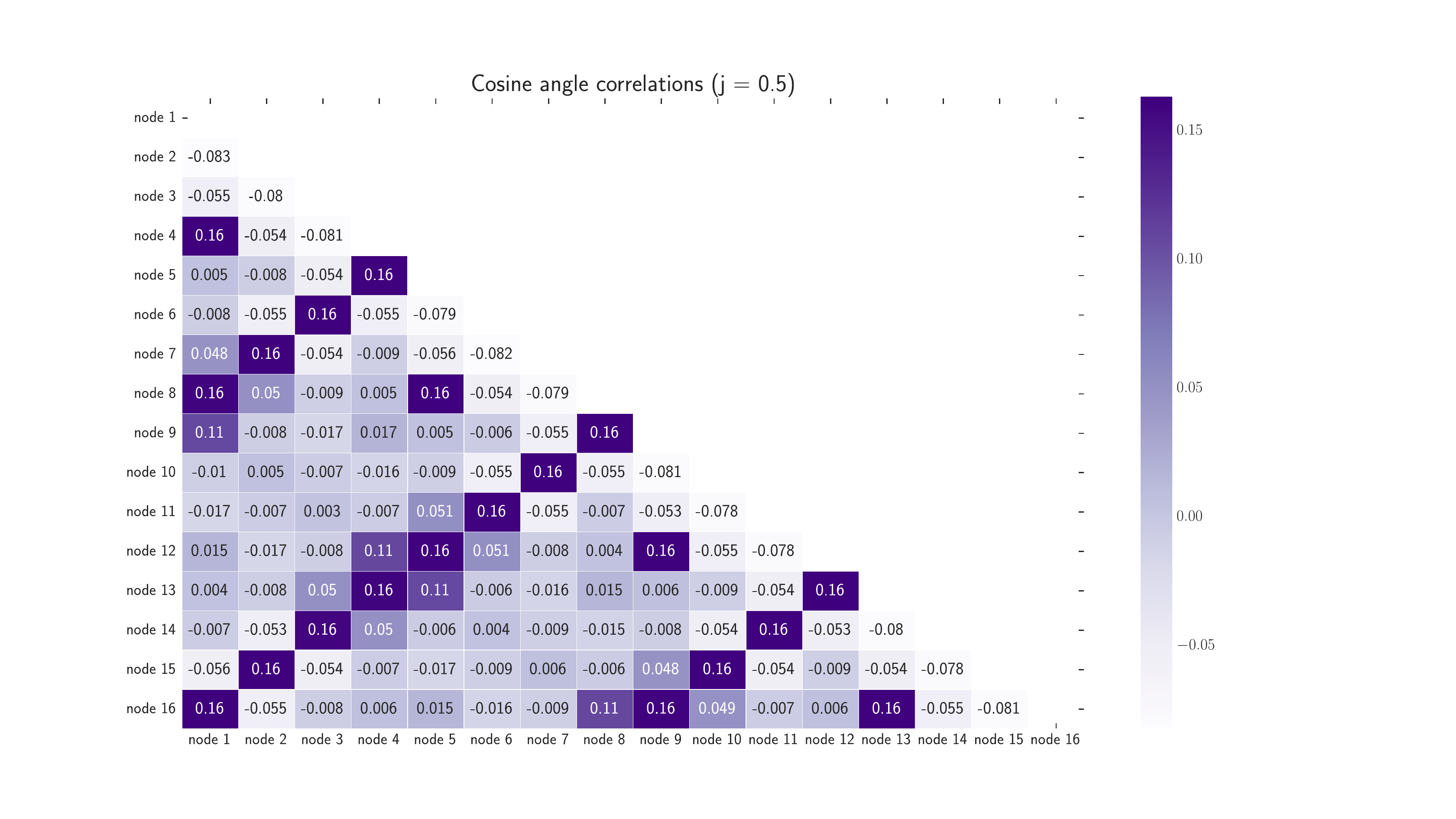}
     \end{subfigure}
     \begin{subfigure}[b]{8.2cm}
     \includegraphics[width=8.5cm]{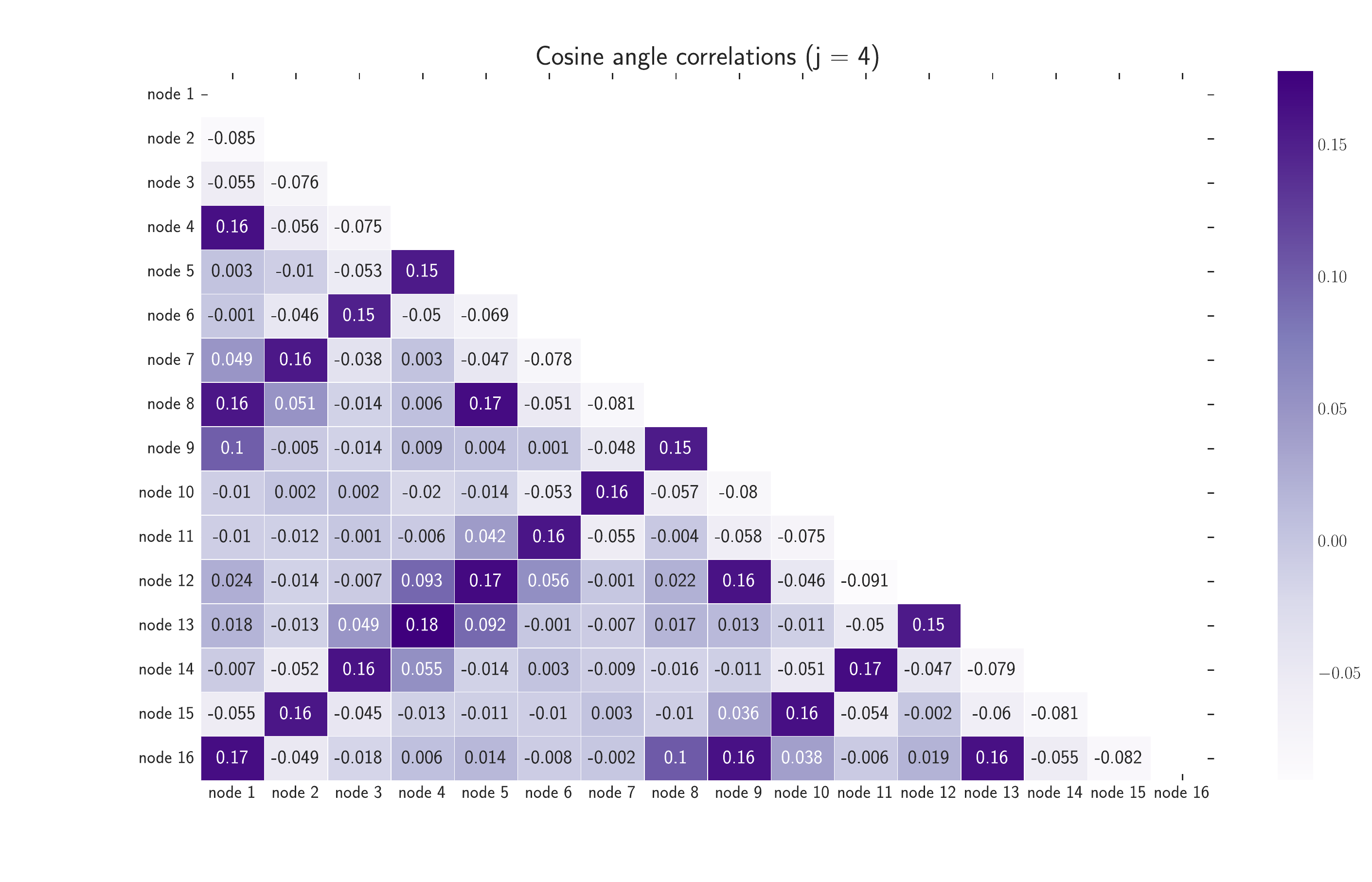}
     \end{subfigure}
    \caption{\label{fig:angles_correlations_table} \textit{Quantum correlations \eqref{correlation_MC_average} computed for all possible couples of nodes at fixed boundary spin $j$. We show explicitly only the minimum and the maximum values $j$. As Figure \ref{fig:angles_correlations} reveals, values in between give similar results. Each node $k$ is equivalent to node $k \pm 4$.} Left: \textit{case $j=0.5$.} Right: \textit{case $j=4$}.}   
\end{figure}

We compute the correlations \eqref{eq:correlations} of the dihedral angle operator \eqref{eq:geom-angleformula} for the 16-cell geometry described in Section \ref{sec:16-cell_triangulation}. We report the result of \eqref{correlation_MC_average} in the right panel of Figure \ref{fig:angles_correlations} and in Figure \ref{fig:angles_correlations_table}. In the left panel of Figure \ref{fig:angles_correlations}, we report the exact same boundary of the 16-cell spinfoam defined in Section \ref{sec:16-cell_triangulation}, labeling the nodes with numbers. We do this to identify the correlations within the 16-cell geometry easily. Looking at the correlation values, we notice that each node $k$ is equivalent to node $k \pm 4$. This is because the 16-cell spinfoam boundary is symmetrical for $90$ degree rotations. Interestingly, there are relatively high values of the correlations between nodes not directly connected by a link (for example, 1-9, 5-13, etc.). In other words, some couples of non-adjacent nodes are strongly correlated. This is the striking difference between the 16-cell geometry and the spinfoam model studied in \cite{MCMC_paper} or the 4-simplex \cite{Gozzini_primordial}. 
In Figure \ref{fig:angles_correlations}, we explicitly show the correlation between just a few couples of nodes not to clutter the picture. We infer that correlations are constant as a function of the boundary spin $j$, confirming the trend observed in \cite{MCMC_paper, Gozzini_primordial}. In Figure \ref{fig:angles_correlations_table}, we show the numerical values of \eqref{correlation_MC_average} computed between all possible combinations to emphasize the complete pattern of nodes. We report the tables for the minimum and maximum values considered for the boundary spins for visualization purposes. From Figure \ref{fig:angles_correlations}, it is evident that the values between $j=0.5$ and $j=4$ have very similar values. 

\newpage
\section{Conclusions}
\label{sec:conclusions}

We have presented a numerical investigation of the non perturbative Hartle-Hawking state defined in covariant loop quantum gravity, in the deep quantum regime. We have computed the mean geometry, as well as fluctuations and correlations in this state.

We have been able to relax the truncation from the tetrahedron triangulation of the cosmological 3-sphere to a 16-cell triangulation.   We have computed the quantities above in the simplified setting where the dynamics is topological.

The numerical analysis is consistent with the hypothesis that refining the triangulation does not affect the global physical picture substantially.   The mean geometry remains consistent with an approximation of a metric threes-sphere.  Fluctuations  and correlations remain high as in the tetrahedral truncation.

We expected a clear dependence of correlation on the separation of the nodes, but we have not found it.   We do not know if this is due to the topological nature of the $BF$ simplification taken. To test so, the calculation must be repeated with the full physical amplitude. This task requires pushing the numerical tools further.

\bigskip\bigskip\bigskip

\begin{center}
{***}
\end{center}
\bigskip\bigskip
\paragraph*{\bf Acknowledgments}
We thank Carlo Rovelli for many discussions on this project and for extensive comments on the final draft of this paper. PF thanks Mikko Karttunen for supporting the development of part of this project within his course in Scientific Computing. PF also warmly thanks Gregory Popovitch for his help in implementing the \texttt{parallel hashmap} for this research project and Hakan T. Johansson for interesting discussions and clarifications about \texttt{fastwigxj}.
We acknowledge the Shared Hierarchical Academic Research Computing Network (SHARCNET) for granting access to their high-performance computing resources. We particularly thank the Compute/Calcul Canada staff for the constant support provided with the Cedar and Graham clusters.
This work was supported by the Natural Science and Engineering Council of Canada (NSERC) through the Discovery Grant "Loop Quantum Gravity: from Computation to Phenomenology." We also acknowledge support from the ID\# 62312 grant from the John Templeton Foundation, as part of the project \href{https://www.templeton.org/grant/the-quantum-information-structure-ofspacetime-qiss-second-phase}{``The Quantum Information Structure of Spacetime'' (QISS)}. The Canada Research Chairs Program at Western University supports FV's research.  The Canada Research Chairs Program supports FV's research at Western University.
FV is affiliated with the Perimeter Institute for Theoretical Physics. The Government of Canada supports research at Perimeter Institute through
Industry Canada and by the Province of Ontario through the Ministry of Economic Development and Innovation.
Western University and Perimeter Institute are located in the traditional lands of Anishinaabek, Haudenosaunee, L\=unaap\`eewak, Attawandaron, and Neutral peoples. %
\bigskip
\bigskip

\appendix

\newpage

\section{Metropolis-Hastings algorithm}
\label{app:M-H_review}
The Metropolis-Hastings algorithm is a Markov Chain Monte Carlo (MCMC) technique used to generate samples from a complex or high-dimensional target probability distribution. 
To discuss the algorithm, we consider a multidimensional discrete variable $x$ on a state space $\chi$, and a quantity $O$ which can be computed as:
\begin{equation}
\label{eq_app:operator_to_compute}
O = \frac{ \sum\limits_{ x \in \chi} W^2_{\chi} \left( x \right) o \left( x \right) }{\sum\limits_{x\in \chi} W^2_{\chi}\left( x \right)} \ ,
\end{equation} 
where $W^2_{\chi}(x)$ is the unnormalized target distribution on $\chi$. We refer to the normalized target distribution as follows:
\begin{equation}
\label{eq_app:normalized_target}
f_{\chi}(x) \equiv \frac{W^2_{\chi} \left( x \right)}{\sum\limits_{x \in \chi} W^2_{\chi}\left( x \right)} \ ,
\end{equation} 
Suppose $W_{\chi}(x)$ can be computed (up to a multiplying constant). In that case, the Metropolis-Hastings algorithm allows constructing an ergodic Markov chain on the space $\chi$ with length $N_{mc}$: $[x]_{1}, [x]_{2}, \ \dots, [x]_{N_{mc}}$ such that each state is indirectly sampled from the normalized target distribution \eqref{eq_app:normalized_target}. We use the notation $[x]_s$ to denote the $s$-th state along the chain. A positive proposal distribution $g_{\chi}$ on the space $\chi$ is required to transit from each state to the next one. 

\medskip

We use the random walk Metropolis-Hastings, in which the sampler locally explores the neighborhood of the current state $[x]_s$ of the Markov chain, proposing a candidate state $[x]'$ sampling from a chosen probability distribution (usually a uniform or a normal distribution). That is, given the current state $[x]_s$, the algorithm suggests a candidate state depending on $[x]_s$. As proposal distribution, in this paper, we consider a truncated normal distribution rounded to integers. Before writing its expression, we first define the cumulative distribution function of a one-dimensional normal distribution with zero mean and standard deviation $\sigma$:  
\begin{equation}
\Phi_{\sigma} (x) = \frac{1}{\sigma \sqrt{2 \pi}} \int_{- \infty}^{x} e^{-\frac{t^2}{2 \sigma^2}} dt \ .
\end{equation}
If the space $\chi$ is multidimensional, $g_{\chi}$ is a multivariate distribution. We sample each component of $x$  from a one-dimensional distribution independent of the others. We write the expression of a one-dimensional normal distribution rounded to integers, truncated between $n_1$ and $n_2$, with standard deviation $\sigma$ as:
\begin{equation}
\label{eq:proposal_explicit}
g_{n_1 , n_2, \sigma} \left( n \right) = \frac{\Phi_{\sigma} (n + 0.5) - \Phi_{\sigma} (n - 0.5)}{\sum\limits_{k = n_1}^{n_2} \left[ \Phi_{\sigma} (k + 0.5) - \Phi_{\sigma} (k - 0.5) \right]} \ .
\end{equation}
For convenience, let's also define the truncated coefficients:
\begin{equation}
\label{eq:truncated_coeff}
C_{n_1 , n_2, \sigma} \left( [x] \right) = \prod\limits_{i=1}^{N} \left \{  \sum\limits_{n=n_1-x_i}^{n_2-x_i} \left[ \Phi_{\sigma} (n + 0.5) - \Phi_{\sigma} (n - 0.5) \right] \right \}\ ,
\end{equation}
where we defined with $x_i$ the $i$-th component of the draw $[x]$. To build the Markov Chain, we compute the ratio between the target distribution \eqref{eq_app:normalized_target} times the truncated coefficients \eqref{eq:truncated_coeff} at the proposal state and the same quantity at the current state. Then, we accept the proposal state with a probability equal to this state. Otherwise, we stay at the current point. The initial steps of the algorithm are usually removed as burn-in iterations during the thermalization phase. The detailed implementation of the Metropolis-Hastings algorithm applied to spinfoams is described in Section \ref{subsec:computing_exp_values_MCMC} of the main text. 

\medskip

After storing a Markov chain with the desired length $N_{mc}$, we can compute \eqref{eq_app:operator_to_compute} applying Monte Carlo on the multidimensional sum over $x$, using the chain itself as a statistical sample. When the number of samples $N_{mc}$ is large enough, we obtain a reasonably good estimate of the original quantity: 
\begin{equation}
\label{eq:app_O_NMC}
O_{{mc}} = \frac{1}{N_{mc}} \sum\limits_{s = 1}^{N_{mc}} o \left([x]_s \right) \approx O \hspace{3mm} \textrm{for $N_{mc} \gg 1$} \ .
\end{equation}
The Monte Carlo estimate \eqref{eq:app_O_NMC} is exactly equal to \eqref{eq_app:operator_to_compute} only in the (ideal) limit of an infinite number of samples. The convergence is faster with respect to the standard version of Monte Carlo \cite{VR_paper} (in which the draws are sampled randomly) because each draw is generated from the distribution \eqref{eq_app:normalized_target}. 

\medskip

There is a non-zero correlation between $[x]_s$ and $[x]_{s+d}$ where $d \geq 1$. This is because each proposed state depends on the previous one (as the process is Markovian). For each quantity \eqref{eq:app_O_NMC} we can compute the autocorrelation function with lag $d$: 
\begin{equation}
\label{eq:autocorrelation}
R_{O} \left( d \right) = \frac{\sum\limits_{s=d+1}^{N_{mc}} \left( o([x]_s) - O_{N_{mc}} \right) \left(  o([x]_{s-d}) - O_{N_{mc}} \right)}{\sum\limits_{s=1}^{N_{mc}} \left( o([x]_s) - O_{N_{mc}} \right)^2} \ .
\end{equation}
Since the Markov Chain converges to a stationary distribution, the autocorrelation \ref{eq:autocorrelation} should decrease as the lag $d$ increases.  

\medskip

Finally, we report in table \ref{tbl:MH_data} the parameters used in the MCMC algorithm \ref{numericalcode} for the calculations considered in this paper.
\begin{table}[H]
\begin{center}
\begin{tabular}{|p{1cm}|p{1cm}|p{1cm}|p{1cm}|p{1cm}|p{1cm}||}
 \hline
 \multicolumn{5}{|c|}{MH - parameters} \\
 \hline
 $j$ & $N_{mc}$ & b & $\sigma$ & C \\
 \hline
 0.5   & $10^6$ & $10$ & $0.40$ & 30 \\
 1.0   & $10^6$ & $10$ & $0.35$ & 30 \\
 1.5   & $10^6$ & $10$ & $0.35$ & 30 \\
 2.0   & $10^6$ & $10$ & $0.35$ & 30 \\
 2.5   & $10^6$ & $10$ & $0.35$ & 30 \\
 3.0   & $10^6$ & $10$ & $0.35$ & 30 \\ 
 3.5   & $10^6$ & $10$ & $0.32$ & 30 \\ 
 4.0   & $10^6$ & $10$ & $0.30$ & 30 \\ 
 \hline
\end{tabular}
\end{center}
\caption{\textit{Parameters used in the MCMC algorithm \ref{numericalcode}. From left to right: \textit{$j$ is the boundary spin of spinfoam amplitude, $N_{mc}$ is the number of iterations over the chain, ``b" is the number of burn-in iterations, $\sigma$ is the standard deviation of the proposal distribution and ``C" corresponds to the number of Markov chains that we stored.}}}
 \label{tbl:MH_data}
\end{table}

\newpage

\section{$SU(2)$ recoupling coefficients}
\label{app:SU(2)_coefficients}
In this Appendix, we report some definitions of standard invariant Wigner symbols. We refer to \cite{GraphMethods, book:varshalovic} for the analytical definitions and graphical representations of the general expressions which appear in the recoupling theory of $SU(2)$ representations. 

\medskip

We explicitly define the $4jm$ symbol, which is less common than the other Wigner symbols usually considered in the literature. It can be obtained by contracting two $3j$ Wigner symbols using an intertwiner $i$. Three inequivalent recouplings correspond to how four spins of $SU(2)$ representations can be coupled in pairs. Each different recoupling corresponds to a (different) $4jm$ symbol for the given set of spins $j_1,j_2,j_3,j_4$. We choose to couple spins $j_1, j_2$ with $j_3, j_4$:
\begin{equation}
\label{eq:4jm_symbol}
\Wfour{ j_1}{ j_2}{ j_3} {j_4}{m_1}{m_2}{m_3}{m_4}{i}\equiv \sum_{m_i} (-1)^{i-m_i} \Wthree{ j_1}{ j_2}{i} {m_1}{m_2}{m_i} \Wthree{ i}{ j_3}{ j_4} {-m_i}{m_3}{m_4}\ .
\end{equation}
The $4jm$ Wigner symbol \eqref{eq:4jm_symbol} has the following orthogonality property:
\begin{equation}
\sum_{m_1,m_2,m_3} \Wfour{ j_1}{ j_2}{ j_3} {j_4}{m_1}{m_2}{m_3}{m_4}{i} \Wfour{ j_1}{ j_2}{ j_3} {j_4}{m_1}{m_2}{m_3}{m_4}{i'} = \frac{\delta_{i, i'}}{d_i} \frac{\delta_{j_{4}, l_{4}}\delta_{m_4, n_4}}{d_{j_4}} \ ,
\end{equation}
Where the triangular inequality constrains the intertwiner $i$:
\begin{equation}
  \max\left\lbrace |j_1-j_2|,|j_3-j_4| \right\rbrace \leq i \leq \min\left\lbrace j_1+j_2,j_3+j_4 \right\rbrace .
\end{equation}
To perform $SU(2)$ integrations and switch to intertwiners on spinfoam boundaries, we use the following property of the 4jm Wigner symbol:  
\begin{equation}
  \label{eq:from_box_to_intertwiners}
 \int_{SU(2)} du \ D^{j_1}_{n_1 m_1}(u)D^{j_2}_{m_2 n_2}(u)D^{j_3}_{m_3 n_3}(u)D^{j_4}_{m_4 n_4}(u)  = \sum\limits_{i} d_i \Wfour{j_1}{j_2}{j_3}{j_4}{n_1}{n_2}{n_3}{n_4}{i} \Wfour{j_1}{j_2}{j_3}{j_4}{m_1}{m_2}{m_3}{m_4}{i} \ ,
\end{equation}
where $D^{j}_{n m}$ represents an $SU(2)$ Wigner matrix. We use the in-line notation for the 6j Wigner symbols
\begin{equation}
\label{eq:6jsymbol}
\{ 6j \} (j_1,j_2,j_3,j_4,j_5,j_6) = \Wsix{j_1}{j_2}{j_3}{j_4}{j_5}{j_6} 
\end{equation}
And the 9j Wigner symbol:
\begin{equation}
\label{eq:9jsymbol}
\{ 9j \} (j_1,j_2,j_3,j_4,j_5,j_6,j_7,j_8,j_9) = \Wnine{j_1}{j_2}{j_3}{j_4}{j_5}{j_6}{j_7}{j_8}{j_9} \ .
\end{equation}
Finally, we report the definition of the 21j Wigner symbol split into 6j and 9j Wigner symbols: 
\begin{align}
\label{eq:21j_exp}
\{ 21j \} &  \left(  j, i_1, i_2, i_3, i_4, b_1, b_2, b_3, p_1, p_2 \right)  = \sum\limits_{g_1} \sum\limits_{g_2} \Big[ \sum\limits_{l} \{ 6j \} \left( j, i_3, j, j, i_4, l \right) \cdot \{ 6j \} \left( i_4, j, l, b_1, j, j \right) \cdot  \\
& \cdot \{ 6j \} \left( l, j, g_2, b_2, j, b_1 \right) \cdot \{ 9j \} \left( l, j, i_3, j, i_2, j, g_2, g_1, j \right) \cdot d_{l} \cdot (-1)^{2l + i_3 + 3 i_4 + b_1 + b_2 + g_2} \Big] \cdot \nn \\ 
& \cdot \{ 6j \} \left( i_2, j, j, i_1, j, g_1 \right) \cdot \{ 6j \} \left( j, p_1, j, j, i_1, g_1 \right) \cdot \{ 6j \} \left( j, b_3, p_2, j, g_2, b_2 \right)  \cdot \{ 6j \} \left( j, p_2, p_1, j, g_1, g_2 \right) \cdot \nn \\
& \cdot d_{g_1} d_{g_2} \cdot (-1)^{2 g_1 + 3i_1 + j + i_2 + 3 g_2 + 2b_2 + b_3 + 2p_1} \nn \ .
\end{align}

\newpage

\begin{thebibliography}{10}


\bibitem{Gozzini_primordial}
F.~Gozzini and F.~Vidotto, ``Primordial fluctuations from quantum gravity,''
{{\em Frontiers in  Astronomy and Space Sciences} {\bfseries 7} (Feb, 2021) }
  

\bibitem{Lehners:2023yrj}
J.~L.~Lehners,
``Review of the no-boundary wave function,''
Phys. Rept. \textbf{1022}, 1-82 (2023)

\bibitem{Bianchi:2010zs}
E.~Bianchi, C.~Rovelli and F.~Vidotto,
``Towards Spinfoam Cosmology,''
Phys. Rev. D \textbf{82}, 084035 (2010)
  
\bibitem{Vidotto:2011qa}
F.~Vidotto,
``Many-nodes/many-links spinfoam: the homogeneous and isotropic case,''
Class. Quant. Grav. \textbf{28}, 245005 (2011)

\bibitem{MCMC_paper}
P.~Frisoni, F.~Gozzini, and F.~Vidotto, ``Markov Chain Monte Carlo methods for
  graph refinement in Spinfoam Cosmology,'' 2022.

\bibitem{libro}
C.~Rovelli and F.~Vidotto,
``Covariant Loop Quantum Gravity: An Elementary Introduction to Quantum Gravity and Spinfoam Theory,''
Cambridge University Press, 2014

\bibitem{Alex}
A.~Perez,
``The Spin Foam Approach to Quantum Gravity,''
Living Rev. Rel. \textbf{16}, 3 (2013)

\bibitem{self_energy_paper}
P.~Frisoni, F.~Gozzini, and F.~Vidotto, ``{Numerical analysis of the
  self-energy in covariant loop quantum gravity},''
 {{\em Phys. Rev. D}
  {\bfseries 105} no.~10, (2022) 106018}

\bibitem{Dona_Frisoni_Ed_infrared}
P.~Don\`a, P.~Frisoni, and E.~Wilson-Ewing, ``{Radiative corrections to the
  Lorentzian Engle-Pereira-Rovelli-Livine spin foam propagator},''
{{\em Phys. Rev. D}  {\bfseries 106} no.~6, (2022) 066022}

\bibitem{Frisoni_2023}
P.~Frisoni, 
{``Studying the
  {EPRL} spinfoam self-energy,''} in {\em The Sixteenth Marcel Grossmann  Meeting}.
\newblock {WORLD} {SCIENTIFIC}, Jan, 2023.
\newblock \url{https://doi.org/10.1142%2F9789811269776_0336}.

\bibitem{VR_paper}
P.~Dona and P.~Frisoni, ``Summing bulk quantum numbers with Monte Carlo in spin
  foam theories,'' 2023.

\bibitem{fastwigxj_related}
J.~Rasch and A.~C.~H. Yu, ``Efficient Storage Scheme for Precalculated Wigner
  3j, 6j and Gaunt Coefficients,''
  \href{http://dx.doi.org/10.1137/S1064827503422932}{{\em SIAM Journal on
  Scientific Computing} {\bfseries 25} no.~4, (2004) 1416--1428}.

\bibitem{Wigxjpf_library}
H.~T. Johansson and C.~Forssén, ``Fast and Accurate Evaluation of Wigner 3$j$,
  6$j$, and 9$j$ Symbols Using Prime Factorization and Multiword Integer
  Arithmetic,'' \href{http://dx.doi.org/10.1137/15m1021908}{{\em SIAM Journal
  on Scientific Computing} {\bfseries 38} no.~1, (Jan, 2016) A376–A384}.
  \url{http://dx.doi.org/10.1137/15M1021908}.

\bibitem{parallel_hash_map}
``Parallel hashmap public repository.''
\newblock \url{https://greg7mdp.github.io/parallel-hashmap/}.

\bibitem{MH_original_paper}
W.~K. Hastings, ``Monte Carlo Sampling Methods Using Markov Chains and Their
  Applications,'' {\em Biometrika} {\bfseries 57} no.~1, (1970) 97--109.
  \url{http://www.jstor.org/stable/2334940}.

\bibitem{MCMC_spinfoam_cosmology}
P.~Frisoni, F.~Gozzini, and F.~Vidotto, ``Markov Chain Monte Carlo methods for
  graph refinement in Spinfoam Cosmology,'' 2022.
\newblock \url{https://arxiv.org/abs/2207.02881}.

\bibitem{GFL_Net_paper}
E.~Bengio, M.~Jain, M.~Korablyov, D.~Precup, and Y.~Bengio, ``Flow Network
  based Generative Models for Non-Iterative Diverse Candidate Generation,''
  2021.
\newblock \url{https://arxiv.org/abs/2106.04399}.

\bibitem{GFL_Net_paper_2}
Y.~Bengio, S.~Lahlou, T.~Deleu, E.~J. Hu, M.~Tiwari, and E.~Bengio, ``GFlowNet
  Foundations,'' 2021.
\newblock \url{https://arxiv.org/abs/2111.09266}.

\bibitem{chandra2001parallel}
R.~Chandra, L.~Dagum, D.~Kohr, R.~Menon, D.~Maydan, and J.~McDonald, {\em
  Parallel programming in OpenMP}.
\newblock Morgan kaufmann, 2001.

\bibitem{harris2020array}
C.~R. Harris, K.~J. Millman, S.~J. van~der Walt, R.~Gommers, P.~Virtanen,
  D.~Cournapeau, E.~Wieser, J.~Taylor, S.~Berg, N.~J. Smith, R.~Kern, M.~Picus,
  S.~Hoyer, M.~H. van Kerkwijk, M.~Brett, A.~Haldane, J.~F. del R{\'{i}}o,
  M.~Wiebe, P.~Peterson, P.~G{\'{e}}rard-Marchant, K.~Sheppard, T.~Reddy,
  W.~Weckesser, H.~Abbasi, C.~Gohlke, and T.~E. Oliphant, ``Array programming
  with {NumPy},'' \href{http://dx.doi.org/10.1038/s41586-020-2649-2}{{\em
  Nature} {\bfseries 585} no.~7825, (Sept., 2020) 357--362}.
  \url{https://doi.org/10.1038/s41586-020-2649-2}.

\bibitem{mckinney2010data}
W.~McKinney {\em et~al.}, ``Data structures for statistical computing in
  python,'' in {\em Proceedings of the 9th Python in Science Conference},
  vol.~445, pp.~51--56, Austin, TX.
\newblock 2010.

\bibitem{GraphMethods}
A.~P. Yutsin, I.~B. Levinson, and V.~V. Vanagas, {\em {Mathematical Apparatus
  of the Theory of Angular Momentum}}.
\newblock Israel Program for Scientific Translation, Jerusalem, Israel, 1962.

\bibitem{book:varshalovic}
V.~D. Aleksandroviic, A.~N. Moskalev, and K.~V. Kel'manoviic, {\em Quantum
  theory of angular momentum: irreducible tensors, spherical harmonics, vector
  coupling coefficients, 3nj symbols}.
\newblock World scientific, 1988.

\end{thebibliography}

\providecommand{\href}[2]{#2}\begingroup\raggedright\endgroup

\end{document}